\documentclass[nofootinbib, twocolumn, pre, aps, 10pt]{revtex4-2} 
\usepackage[utf8]{inputenc}
\usepackage{longtable}
\usepackage{xcolor}
\usepackage{blindtext,alltt}
\usepackage{amsmath}
\usepackage{mathrsfs}
\usepackage{amsfonts}
\usepackage{amssymb}
\usepackage{graphicx}
\usepackage{mathrsfs}
\usepackage{hyperref}

\newcommand{\e}{\textrm{e}}

\newcommand{\bra}[1]{\big \langle #1 \big|}
\newcommand{\ket}[1]{\big | #1 \big \rangle}
\newcommand{\braket}[2]{\big \langle #1 \big | #2 \big \rangle}
\newcommand{\melem}[3]{\big \langle #1 \big | #2 \big| #3 \big \rangle}
\newcommand{\Ham}{\widehat{\mathrm{H}}}
\newcommand{\MC}{\, \overset{\textrm{M.C.}}{=}\, }
\newcommand{\jsum}[3]{\,\,\underset{\hspace{-0.75em}{\scriptstyle #1}}{\sum\nolimits_{#2}^{#3}}\,}
\newcommand{\Pv}{\mathchoice{\raisebox{0.2em}{\scalebox{1.3}{$\wp$}}}{\raisebox{0.2em}{\scalebox{1.3}{$\wp$}}}{\raisebox{0.1em}{\scalebox{0.9}{$\wp$}}}{\raisebox{0.1em}{\scalebox{0.8}{$\wp$}}}}
\newcommand{\Pvp}{\Pv^{\prime}}
\newcommand{\Pvpp}{\Pv^{\prime \prime}}
\newcommand{\Pvb}{\bar{\Pv}}
\newcommand{\Pvt}{\widetilde{\Pv}}
\newcommand{\Pvh}{\widehat{\Pv}}
\newcommand{\U}{\mathcal{U}}

\begin{document}
	
\title{Monte Carlo generation of localised particle trajectories}

\author{Ivan Ahumada}
\email{ivan.ahumada@umich.mx (Corresponding Author)}
\affiliation{Instituto de F\'isica y Matem\'aticas,\\
Universidad Michoacana de San Nicol\'as de Hidalgo, Morelia, M\'exico.}

\author{James P. Edwards}
\email{james.p.edwards@plymouth.ac.uk}
\affiliation{Centre for Mathematical Sciences, University of Plymouth, Plymouth, PL4 8AA, UK}

\begin{abstract}
	We introduce modifications to Monte Carlo simulations of the Feynman path integral that improve sampling of localised interactions. The new algorithms generate trajectories in simple background potentials designed to concentrate them about the interaction region, reminiscent of importance sampling. This improves statistical sampling of the system and overcomes a long-time ``undersampling problem'' caused by the spatial diffusion inherent in Brownian motion. We prove the validity of our approach using previous analytic work on the distribution of values of the Wilson line over path integral trajectories and illustrate the improvements on some simple quantum mechanical systems.
\end{abstract}

\maketitle	

\section{Introduction}
The (Feynman) propagator is of fundamental importance in quantum mechanics, as the integral kernel of the time evolution operator, $\,\widehat{\U}(T)$, in configuration space. Expressing this operator in the position basis, $\big\{\ket{x}\big\}$ normalised to $\braket{x}{y} = \delta^{D}(x-y)$, 
\begin{equation}
	\widehat{\U}(T) = \int d^{D}x\int d^{D}y\,  K(y, x; T) \ket{y}\bra{x}\, ,
\end{equation}
introduces the kernel as the matrix elements 
\begin{equation}
	\hspace{-1.6em}	K(y, x; T) := \melem{y}{\widehat{\U}(T)}{x} \longrightarrow \melem{y}{\e^{-i\Ham T}}{x}\,, \quad (T \geqslant 0)\, 
\end{equation}
(assuming a static Hamiltonian, $\Ham$). As is well-known, obtaining the kernel is equivalent to solving the system, yet this can be analytically challenging unless the system is particularly simple or enjoys special symmetries. 

This article presents improved numerical algorithms for estimating the kernel using the ``Worldline Monte Carlo'' (WMC) technique. In the imaginary time formalism $K(y,x;T)$ can be found from the path integral over trajectories propagating from $x$ to $y$ in time $T$:
\begin{equation}
	K(y,x; T) = \int_{x(0) = x}^{x(T) = y} \hspace{-1.5em} \mathscr{D}x(\tau) \, \e^{-\int_{0}^{T} d\tau \big[ \frac{m\dot{x}^{2}}{2} + V(x(\tau))  \big]} \,,
	\label{eqKPathInt}
\end{equation}
with $V(x)$ the potential defined canonically according to $\Ham = \frac{\widehat{p}^{2}}{2m} + V(\widehat{x})$. The WMC approach goes back to early work on bound states in \cite{TjonThesis,Nieuwenhuis:1995ux, Nieuwenhuis:1996mc, Savkli:1999rw, Savkli:2002fj}, but its adaptation \cite{Gies:2001zp, Gies:2001tj,  Schmidt:2002mt, Langfeld:2002vy, Gies:2003cv, Gies:2005sb, Langfeld:2007wh} to simulations based on the worldline formalism of quantum field theory \cite{Strass1, SchmidtRev, ChrisRev, UsRep} and to processes in background fields \cite{Gies:2005bz, Gies:2005ym}, the Casimir effect \cite{Moyaerts:2003ts, Gies:2006cq, Gies:2006bt, Weber:2009dp, PhysRevLett.102.060402, Idrish, Idrish2} and propagators in flat and curved space \cite{UsMonteCarlo, Franchino-Vinas:2019udt, Corradini:2020tgk}, as well as fermionic models \cite{dunne2009worldline}, has established a powerful and universal approach to estimating path integrals.

However, as discussed below, WMC suffers a late time (i.e. large $T$) loss of precision. This  ``undersampling'' problem, caused by the $\sqrt{T}$ spatial growth of Brownian motion trajectories, leads to poor sampling of localised potentials. This has been a limiting factor for precise estimations of the propagator. Here we modify the WMC algorithms that generate trajectories to control this diffusion, concentrating them about the support of $V(x)$. 

Naturally this causes a bias in the WMC simulations, as trajectories no longer diffuse correctly. We provide two methods (analytic and numerical) to remove this bias by modifying the Gaussian weight on particle \textit{velocities}. We not only recover the desired sampling of the quantum system, but simulations based on this approach no longer suffer from undersampling and hence improve the large time estimation of the kernel, since they better sample the potential. We immediately obtain order of magnitude improvements in estimations of ground state energies.

The two methods can be summarised as follows. Denoting $v \equiv v[x] := \int_{0}^{T} V(x(\tau)) d\tau$ (the integral of the potential along the trajectory $x$), the WMC simulation estimates the expectation value of this Wilson line from $N_{L}$ trajectories ($\MC$ indicates Monte Carlo estimate):
\begin{equation}\vspace{-0.5em}
	K(y, x; T) \MC K_{0}(y,x;T)\frac{1}{N_{L}}\jsum{v_{i}\sim \Pv(v)}{i=1}{N_{L}} \e^{-v_{i}}\,,
	\label{eqKMC}
\end{equation}
where the samples, $v_{i}$, follow the distribution $\Pv(v | y, x; T)$ described below, inherited from the free particle Gaussian weight on velocities \cite{PvHz}, and $K_{0}$ is the free particle kernel. Instead we generate trajectories in a background potential, $U$, with distribution on the $\{v_{i}\}$ now given by $\Pv_{U}$. Our analytic approach relies on a \textit{compensating factor}, $F(v) \equiv \frac{\Pv(v)}{\Pv_{U}(v)}$ with which
\begin{equation}
	K(y, x; T) \MC K_{0}(y,x;T)\frac{1}{N_{L}}\jsum{v^{U}_{i}\sim \Pv_{U}(v)}{i=1}{N_{L}} \frac{\e^{-v_{i}^{U}}}{F(v_{i}^{U})}
	\label{eqKmcCompensate}
\end{equation}
is a correct Monte Carlo estimation of the kernel when the $\{v_{i}^{U}\}$ follow $\Pv_{U}(v)$. 

Finding the compensating factor is non-trivial in general\footnote{Section \ref{secPAP} derives the distributions as integral transforms of a kernel: their determination equates to solving the system \cite{PvHz}.}, so in the second, universally applicable numerical approach, the background can be compensated by a \textit{potential subtraction}, whereby the Wilson line variable becomes $\nu\equiv \nu[x] := \int_{0}^{T} d\tau\big[ V(x(\tau)) - U(x(\tau))\big]$, with values distributed as $\nu \sim \Pv_{\bar{U}}(\nu)$. We will prove equivalence of 
\begin{equation}
	K(y, x; T) \MC K_{U}(y,x;T)\frac{1}{N_{L}}\jsum{\nu_{i}\sim \Pv_{\bar{U}}(\nu)}{i=1}{N_{L}} \e^{-\nu_{i}}\,,
\end{equation}
where $K_{U}$ is the kernel in the potential $U$ alone. 

We begin in section \ref{secWLN} with an outline of the WMC approach, emphasising the late time undersampling problem. We then analyse the distributions on the Wilson line variables in section \ref{secWLstats}, leading to our new numerical algorithms (pseudocode is in Appendix \ref{secApAlg}). There and in section \ref{secPot} we describe how to compensate for these modifications. We illustrate their application in \ref{secApps} by estimating ground state energies for some simple systems. 

\section{Worldline numerics}
\label{secWLN}
The main idea of WMC was proposed in \cite{TjonThesis, Nieuwenhuis:1995ux, Nieuwenhuis:1996mc} and corresponds to replacing the continuous integral over trajectories in (\ref{eqKPathInt}) by a finite sum  $\int \mathscr{D}x(\tau) \rightarrow \frac{1}{N_{L}}\sum_{i = 1}^{N_{L}}$ over $N_{L}$ paths (originally closed ``loops'') $\{x_{i}(\tau)\}_{i=1}^{N_{L}}$. To evaluate the Wilson line, we further discretise these trajectories in $\tau$, so $x_{i}(\tau) \rightarrow \{x_{i}(\tau_{k})\}_{k = 1}^{N_{P}}$ becomes a set of $N_{P}$ points (we do \textit{not} discretise the target space). Numerical implementation rescales to the dimensionless variable $u := \frac{\tau}{T}$ and expands about the straight line between the endpoints using \textit{unit trajectories}, $q(u)$,
\begin{equation}
	x(\tau) = x + (y-x)u + \sqrt{\frac{T}{m}}q(u)\,,
	\label{eqxq}
\end{equation}
thus arriving at (a normalised expectation value, $\langle 1 \rangle = 1$)
\begin{align}
	\frac{K(y,x; T)}{K_{0}(y,x ; T)} &= \Big\langle \e^{-T \int_{0}^{1} du\, V(x(u)) } \Big\rangle  \nonumber \\
	&\!\!\!\MC  \frac{1}{N_{L}}\sum_{i = 1}^{N_{L}}\,\e^{- \frac{T}{N_{P}}\sum_{k=1}^{N_{P}} V(x_{i}(u_{k})) }\,,
	\label{eqKDiscrete}
\end{align}
where the fluctuations $q_{i} = q(u_{i})$ with $u_{i} = \frac{i}{N_{P}}$ should have a Gaussian distribution on velocities
\begin{equation}
	\hspace{-1em}\mathcal{P}[q(u)] \propto \e^{\!- \frac{1}{2}\int_{0}^{1} du\, \dot{q}^{2}} \longrightarrow \e^{-\frac{N_{P}}{2} \sum_{k=1}^{N_{P}} (q_{k} - q_{k-1})^{2}}\,,
	\label{eqPq}
\end{equation}
and satisfy Dirichlet boundary conditions (DBC), $q_{0} = 0 = q_{N_{P}}$. We thereby identify the discretised Wilson line variables as $v_{i} := v[x_{i}] = \frac{T}{N_{P}}\sum_{k=1}^{N_{P}} V(x_{i}(u_{k}))$.

Application of worldline numerics in field theory exploits the first quantised worldline approach based on path integrals over \textit{relativistic} point particles. Indeed, these methods were adapted to non-relativistic quantum mechanics (which we focus on here) only recently, numerically in \cite{UsMonteCarlo, Corradini:2020tgk} and analytically -- for the same type of localised potential studied here -- in \cite{PvHz, nHit} (also \cite{SebaNaser}). 

Trajectories distributed according to (\ref{eqPq}) can be generated directly with various algorithms \cite{Gies:2006cq, Franchino-Vinas:2019udt}. It is important that they are not rejection-based, as Metropolis-type algorithms are. Instead, a direct sampling of the distributions implies that all trajectories generated contribute to the estimation of physical quantities (alternative algorithms based on thermalisation also exist \cite{Nieuwenhuis:1995ux, Nieuwenhuis:1996mc, Gies:2001zp, Gies:2001tj}). Crucially, the modified algorithms proposed in this work preserve this property, in that the effects of generating trajectories in background potentials are ``subtracted'' without any  accept / reject step.

\subsection{Undersampling problem}
\begin{figure}[h]
	\centering
	\includegraphics[width=0.5\textwidth]{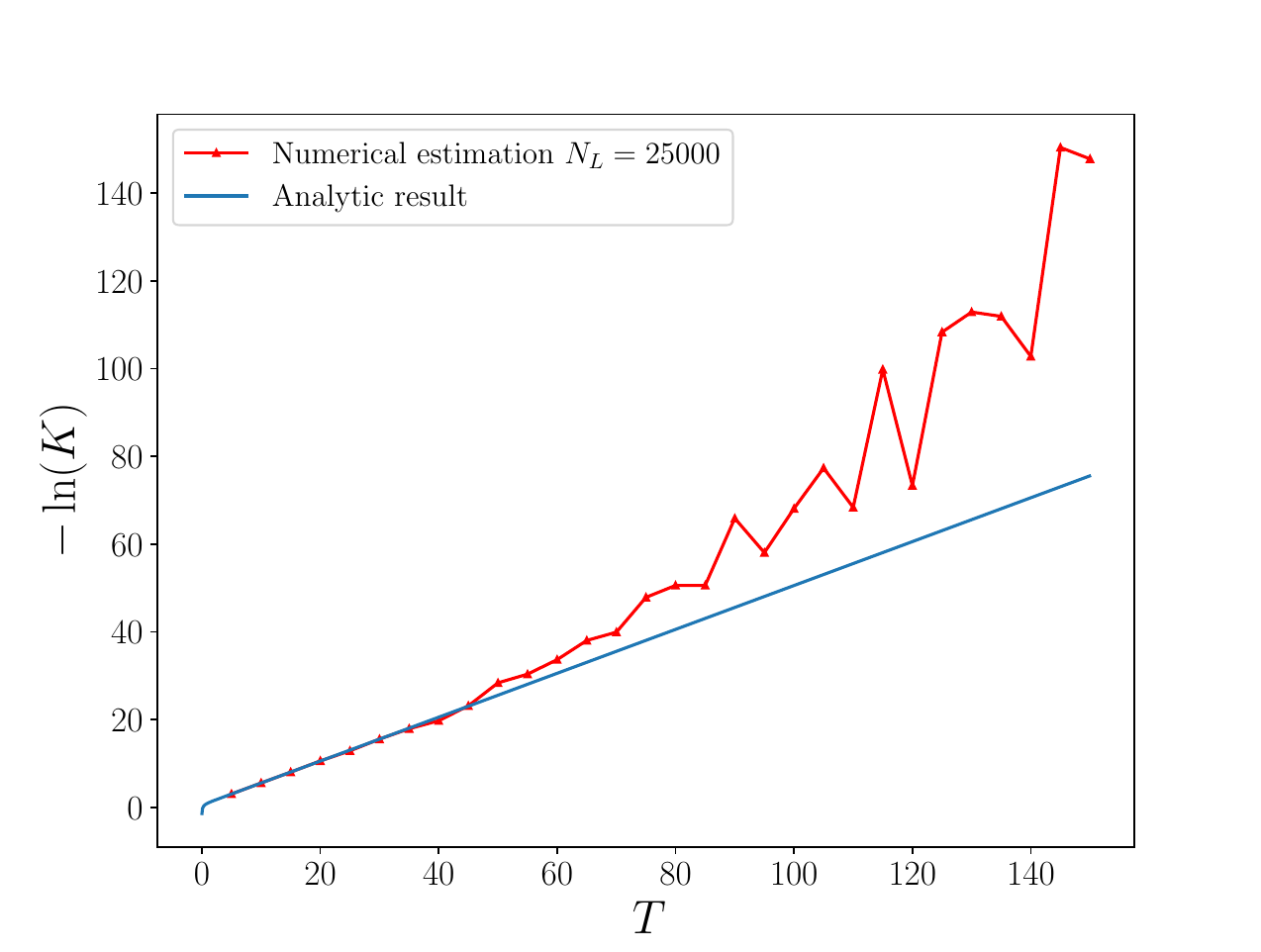}
	\caption{An illustrative WMC estimation of the kernel, $K$, for the harmonic oscillator using $N_{L} = 25000$ trajectories and $N_{p}=5000$ points. Undersampling is manifest in the late time deviation from linearity as predicted by (\ref{eqKlimit}).}
	\label{figUndersampling}
\end{figure}

References \cite{Gies:2005sb, UsMonteCarlo} show that the $\sqrt{T}$ spatial growth of trajectories -- see (\ref{eqxq}) -- causes late time diffusion away from regions that dominate estimation of the potential in (\ref{eqKDiscrete}). For systems with energies bounded from below, the spectral decomposition provides the asymptotics
\begin{equation}
	K(y,x; T) \overset{T \rightarrow \infty}{\sim} \psi_{0}(y)\psi_{0}^{\star}(x)\e^{-T E_{0}}\,,
	\label{eqKlimit}
\end{equation} 
where $E_{0}$ is the ground state energy and $\psi_{n}(x) := \braket{x}{\Psi_{n}}$ are energy eigenfunctions. Undersampling is shown for the harmonic oscillator in Fig \ref{figUndersampling}: \textit{larger} values of $V(x)$ are sampled for large $T$, underestimating $\log(K(y,x; T))$ which deviates markedly from the linearity expected from (\ref{eqKlimit}). This spoils estimation of $E_{0}$ as the asymptotic gradient $E_{0} = -\lim_{T \rightarrow \infty}\frac{\partial}{\partial T}\log(K(y, x; T))$.

One cause of undersampling is that the Gaussian weight on velocities in (\ref{eqPq}) lacks information on the potential, $V(x)$. Here, we propose new algorithms that generate trajectories in background potentials rather than as free particles, tuned to favour trajectories that better sample the system in question. To overcome the bias this induces, we must examine the distribution of the Wilson line variable, $v$, for trajectories following (\ref{eqPq}).

\subsection{Path averaged potential}
\label{secPAP}
In \cite{PvHz}, motivated by \cite{Gies:2005bz}, one of the authors began systematic studies of the statistical distribution on values of the Wilson line variable. This ``Path Averaged Potential'' (PAP) is defined via a constrained path integral,
\begin{equation}
	\hspace{-0.9em}\Pvb(v | y, x; T) :=  \int_{x(0) = x}^{x(T) = y} \hspace{-1.5em} \mathscr{D}x(\tau) \, \delta\Big( v - v[x]) \Big) \e^{-\int_{0}^{T} d\tau \frac{m\dot{x}^{2}}{2}} \,,
	\label{eqPvdef}
\end{equation}
normalised to define $\Pv(v | y, x; T)= \frac{\bar{\Pv}(v | y, x; T)}{K_{0}(y,x;T)}$.

\begin{figure}
	\centering
	\includegraphics[width=0.5\textwidth]{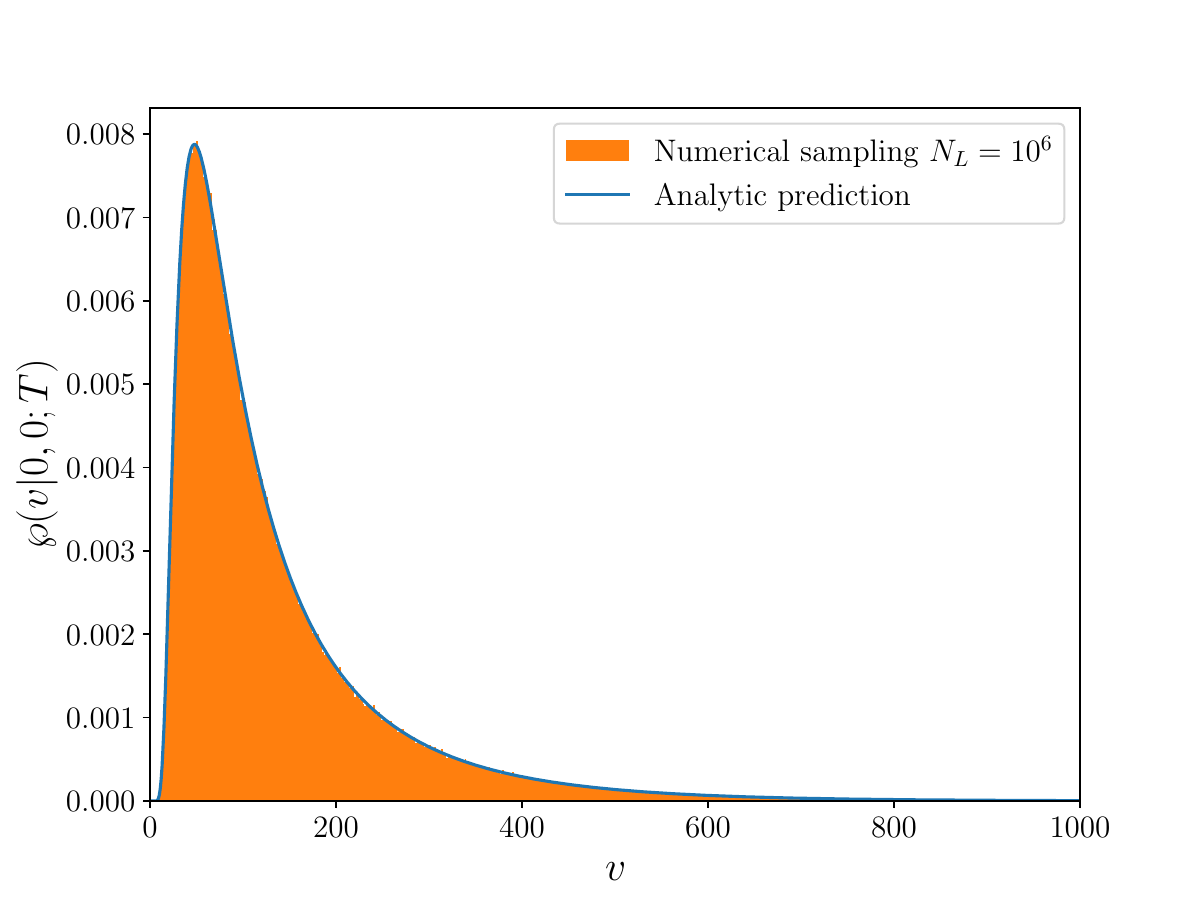}
	\caption{The distribution $\Pv( v | 0, 0, 40)$ for a harmonic oscillator ($m = 1$, $\omega = 1$) sampled with $N_{L} = 10^{6}$ trajectories. The histogram of sampled values of $v[x]$ is shown against the analytic result (blue solid line), with good agreement.}
	\label{figPvHO}
\end{figure}

This distribution on the space of trajectories describes the path integral contribution of paths with a fixed value, $v$, of the Wilson line variable and defines an invertible integral transform of the kernel:
\begin{align}
	K(y,x; T) &= \int_{-\infty}^{\infty} dv\, \bar{\Pv}(v | y, x; T)\e^{-v}\,,\nonumber \\
	\bar{\Pv}(v | y, x; T) &= \frac{1}{2\pi} \int_{-\infty}^{\infty} dz\, \e^{ivz} \widetilde{K}(y, x; T, z)\,,
	\label{eqIntTransf}
\end{align}
where in $\widetilde{K}(y, x; T, z)$ we continue $V(x) \rightarrow izV(x)$. An example of this distribution for the harmonic oscillator is in Figure \ref{figPvHO}, with a numerical sampling using WMC.

Identification of the distribution on $v$ is crucial for the analysis in this work, mapping between configuration space and a complementary ``Wilson line space,'' from which values of $v$ are drawn.

\section{Wilson line statistics}
\label{secWLstats}
Equation (\ref{eqIntTransf}) shows that good Monte Carlo sampling equates with drawing values of $v[x]$ from the appropriate $\Pv(v)$. Moreover, smaller values of $v$ tend to contribute more to the formation of the kernel in (\ref{eqIntTransf}), yet this small-$v$ tail was shown in \cite{UsMonteCarlo} to be sampled poorly.
This suggests a strategy to mitigate undersampling: we developed algorithms that generate trajectories in background potentials, $U(x)$, taken to be a harmonic oscillator, $U_{\Omega}(x) =\frac{1}{2}m \Omega_{i}^{2} x_{i}^{2}$, or a linear potential, $U_{\kappa} = \kappa_{i} x_{i}$,  designed respectively for approximately symmetric potentials and for potentials with spatially skewed dominant features. Thus we change (\ref{eqPq}) to
\begin{equation}
	\mathcal{P}[q(u)] \propto \exp{\Big(-\int_{0}^{1} du \Big[\frac{\dot{q}^{2}}{2} + T^{2}U\Big(\sqrt{\frac{T}{m}}q(u)\Big)\Big]\Big)}
	\label{eqPqU}
\end{equation}
(this is sufficient for $U_{\Omega}$ and $U_{\kappa}$ since boundary terms in $x(\tau)$ can be absorbed by shifting $q(\tau)$). Then, for symmetric, localised potentials, $V(x)$, we superimpose a harmonic oscillator background centered at the potential's minimum, that favours trajectories in this region; for potentials with skewed features (we will call them ``one-side dominated'' potentials), the linear background encourages trajectories towards smaller values of $V(x)$, providing the largest contributions to the path integral.

We must point out that a harmonic background was used in heat kernel simulations on curved space in \cite{Corradini:2020tgk}; however, in contrast to their ``regulating mass''  valid for \textit{small} propagation times, we interpret the modification as a genuine background potential and, crucially, show how to compensate for the bias induced by such backgrounds.

\subsection{Monte Carlo distributions}
We present the newly developed algorithms that produce discretised trajectories, $\{x^{\Omega}_{i}\}$ and $\{x^{\kappa}_{i}\}$, according to (\ref{eqPqU}) in the appendix. In the main text we focus on the statistical distributions followed by the Wilson line variable in the presence of these backgrounds. 

As it stands, a WMC simulation with trajectories generated according to (\ref{eqPqU}) produces samples of the modified Wilson line variable, denoted by $\{v^{\Omega}_{i}\}$ and $\{v^{\kappa}_{i}\}$. These will of course follow different distributions, $v^{\Omega} \sim \Pv_{\Omega}(v^{\Omega})$ and $v^{\kappa} \sim \Pv_{\kappa}(v^{\kappa})$. Such simulations would make incorrect Monte Carlo estimations, respectively
\begin{align}
	\hspace{-1em}K_{0}\Big\langle \e^{-\int_{0}^{T} V(x(\tau)) d\tau} \Big\rangle_{\Omega} &\MC \frac{K_{0}}{N_{L}}\jsum{v_{i}^{\Omega}\sim \Pv_{\Omega}(v)}{i=1}{N_{L}} \e^{-v_{i}^{\Omega}} \\
	&\MC  K_{0}\int_{-\infty}^{\infty}\!\Pv_{\Omega}(v)\, \e^{-v} dv \,,\label{eqSimOmega} \\
	\hspace{-1em}K_{0}\Big\langle \e^{-\int_{0}^{T} V(x(\tau)) d\tau} \Big\rangle_{\kappa} &\MC \frac{K_{0}}{N_{L}}\jsum{v_{i}^{\kappa}\sim \Pv_{\kappa}(v)}{i=1}{N_{L}} \e^{-v_{i}^{\kappa}} \\
	&\MC  K_{0}\int_{-\infty}^{\infty}\!\Pv_{\kappa}(v)\, \e^{-v} dv \,,
	\label{eqSimkappa}
\end{align}
where we suppress unimportant arguments to functions. The notation $\langle \cdots \rangle_{\Omega}$ and $\langle \cdots \rangle_{\kappa}$ indicate expectation values calculated in the appropriate background:
\begin{align}
	\label{eqExpOmegaKappa}
	\langle \cdots \rangle_{\bullet} &:= \frac{1}{K_{\bullet}}\int \mathscr{D}x(\tau)\, \cdots \,\e^{- \int_{0}^{T}d\tau\, \big[ \frac{m\dot{x}^{2}}{2} +U_{\bullet}(x)\big]}\,, 
\end{align}
normalised by the kernel in the presence of the background (i.e with $\cdots \rightarrow 1$). Likewise, the distributions on the values of the Wilson line are accordingly
\begin{align}
	\label{eqPvOmegaKappaDef}
	\Pv_{\bullet}(v) &= \big\langle \delta \big( v - v[x]d\tau \big)\big\rangle_{\bullet}\,.
\end{align}
Hence we must compensate for the effect of the background on the spatial distribution of the trajectories.

From (\ref{eqSimOmega}-\ref{eqSimkappa}) follows an immediate solution: assuming we can identify functions $F(v)$ and $G(v)$ such that $\Pv_{\Omega}(v)\equiv F(v)\Pv(v)$ and $\Pv_{\kappa}(v)\equiv G(v)\Pv(v)$, the following should be faithful Monte Carlo simulations:
\begin{align}
	\label{eqSimOmegaF}
	\hspace{-1em} \frac{K_{0}}{N_{L}}\,\jsum{v_{i}^{\Omega}\sim \Pv_{\Omega}(v)}{i=1}{N_{L}}\frac{\e^{-v_{i}^{\Omega}}}{F(v_{i}^{\Omega})} \MC  K_{0}\int_{-\infty}^{\infty}\Pv(v)\e^{-v}dv = K_{V} \,,\\
	\hspace{-1em} \frac{K_{0}}{N_{L}}\,\jsum{v_{i}^{\kappa}\sim \Pv_{\kappa}(v)}{i=1}{N_{L}} \frac{\e^{-v_{i}^{\kappa}}}{G(v_{i}^{\kappa})} \MC  K_{0}\int_{-\infty}^{\infty}\Pv(v)\e^{-v} dv = K_{V}\,.
	\label{eqSimkappaG}
\end{align}
We prove these claims for harmonic oscillator and linear backgrounds in the following section.

\subsubsection{Analytic distributions}
Here explicit calculations are presented in both background potentials for some simple systems that allow us to justify (\ref{eqSimOmegaF}) and (\ref{eqSimkappaG}). To allow analytic determination of the relevant distributions we treat the harmonic oscillator ($V(x) = \frac{1}{2}m\omega^{2}x^{2}$) in the background $U_{\Omega}$ and the linear potential ($V(x) = kx$) in the background $U_{\kappa}$, in one dimensional quantum mechanics for simplicity.

By emulating the steps in \cite{PvHz}, it is straightforward to obtain the multiplicative relations for these systems\footnote{Explicit formulae for the PAPs of these potentials are given in Appendix \ref{secApPAP}, following \cite{PvHz}.},
\begin{align}
	\label{eqPvOmegaF}
	\Pv_{\Omega}(v | y, x; T) &= \sqrt{\frac{\sinh(\Omega T)}{\Omega T}}\, \e^{-\frac{\Omega^{2}}{\omega^{2}}v}\, \Pv(v | y,x; T)\,,\\
	\Pv_{\kappa}(v | y, x; T) &= \Pv\big(v + \frac{\kappa kT^{3}}{12m}\big| y,x; T\big)\nonumber \\
	&= \e^{\frac{\kappa T}{2}(x+y) - \frac{\kappa^{2}T^{3}}{24}}\, \e^{-\frac{\kappa}{k}v} \, \Pv(v | y,x; T)\,,
	\label{eqPvkappaG}
\end{align}
which identify the \textit{compensation factors} $F(v)=\sqrt{\frac{\sinh(\Omega T)}{\Omega T}}\, \e^{-\frac{\Omega^{2}}{\omega^{2}}v}$ and ${G(v)= \e^{\frac{\kappa T}{2}(x+y)-\frac{\kappa^{2}T^{3}}{24}}}\,\e^{-\frac{\kappa}{k}v} $.

To verify that the compensation factor $F(v_{i}^{\Omega})$ is correct, interpret the modified sum in \eqref{eqSimOmegaF} as a transformation on the $\{v_{i}^{\Omega}\}$. Then it can be written 
\begin{eqnarray}
	\frac{K_{0}}{N_{L}}\jsum{v_{i}^{\Omega}\sim \Pv_{\Omega}(v)}{i=1}{N_{L}} \frac{\e^{-v_{i}^{\Omega}}}{F(v_{i}^{\Omega})}
	&\equiv & \frac{K_{0}}{N_{L}}\jsum{v_{i}^{\prime}\sim \Pvp(v)}{i=1}{N_{L}}\e^{-v_{i}^{\prime}}\,,
	\label{eqSumMCF}
\end{eqnarray}
where we have defined a new set of Wilson line variables $\{v_{i}^{\prime}\}:=\{v_{i}^{\Omega}+\log(F(v_{i}^{\Omega}))\}$. This modified average is a Monte Carlo estimation of the following integral:
\begin{equation}
	\frac{K_{0}}{N_{L}}\jsum{v_{i}^{\prime}\sim \Pvp(v)}{i=1}{N_{L}}\e^{-v_{i}^{\prime}} \MC K_{0}\int_{-\infty}^{\infty}\Pv^{\prime}(v)\e^{-v}dv, \label{eqPvPrimedef}
\end{equation}
with a new distribution, $\Pv^{\prime}$, inherited from $\Pv_{\Omega}$. Elementary probability theory gives
\begin{equation}
	\hspace{-0.5em}	\Pv^{\prime}(v)=\Pv_{\Omega}\Big(\Big(v-\log\sqrt{\frac{\sinh(\Omega T)}{\Omega T}}\Big)\frac{\omega^{2}}{\omega^{2}-\Omega^{2}}\Big)\frac{\omega^{2}}{\omega^{2}-\Omega^{2}}.
	\label{eqPvPrime}
\end{equation}
To see that this is the correct distribution we use it in \eqref{eqPvPrimedef} changing variables to $\big(v-\log\sqrt{\frac{\sinh(\Omega T)}{\Omega T}}\big)\frac{\omega^{2}}{\omega^{2}-\Omega^{2}}$ for
\begin{align}
	\hspace{-0.6em}\frac{K_{0}}{N_{L}}\jsum{v_{i}^{\prime}\sim \Pvp(v)}{i=1}{N_{L}}\e^{-v_{i}^{\prime}} & \MC K_{0}\int_{-\infty}^{\infty}\Pv_{\Omega}(v)\e^{-v}\sqrt{\frac{\Omega T}{\sinh(\Omega T)}}\e^{\frac{\Omega^{2}}{\omega^{2}}}dv \nonumber  \\
	\hspace{-1em}&=K_{0}\int_{-\infty}^{\infty}\Pv(v)\e^{-v}dv=K_{V},
\end{align}
using (\ref{eqPvOmegaF}), as required. 

We verify the compensation factor $G(v_{i}^{\kappa})$ analogously, treating the sum in \eqref{eqSimkappaG} as a transformation on the $\{v_{i}^{\kappa}\}$. So we write the sum as
\begin{eqnarray}
	\frac{K_{0}}{N_{L}}\jsum{v_{i}^{\kappa}\sim \Pv_{\kappa}(v)}{i=1}{N_{L}} \frac{\e^{-v_{i}^{\kappa}}}{G(v_{i}^{\kappa})}
	&\equiv &\frac{K_{0}}{N_{L}}\jsum{v_{i}^{\prime \prime}\sim \Pvpp(v)}{i=1}{N_{L}}\e^{-v_{i}^{\prime\prime}},
	\label{eqSumMCG}
\end{eqnarray}
where now $\{v_{i}^{\prime\prime}\}:=\{v_{i}^{\kappa}+\log(G(v_{i}^{\kappa}))\}$. This average is a Monte Carlo estimation of an integral:
\begin{equation}
	\frac{K_{0}}{N_{L}}\jsum{v_{i}^{\prime \prime}\sim \Pvp(v)}{i=1}{N_{L}}\e^{-v_{i}^{\prime\prime}}\MC 	K_{0}\int_{-\infty}^{\infty}\Pvpp(v)\e^{-v}dv, \label{eqPvPrimePrimedef}
\end{equation}
with a new distribution, $\Pvpp$, induced by $\Pv_{\kappa}$:
\begin{equation}
	\hspace{-1em}\Pvpp(v)=\Pv_{\kappa}\Big(\frac{k}{k-\kappa}\Big[v+\frac{\kappa T}{2}\Big(\frac{\kappa T^2}{12}-(x+y)\Big)\Big]\Big)\left|\frac{k}{k-\kappa}\right|.
	\label{eqPvPPrime}
\end{equation}
Substituting into \eqref{eqPvPrimePrimedef} with a change of variables to $\frac{k}{k-\kappa}\big[v+\frac{\kappa t}{2}(\frac{\kappa t^2}{12}-(x+y))\big]$ and the help of (\ref{eqPvkappaG})
proves that (\ref{eqSumMCG}) recovers a correct estimation of the propagator. 
\subsubsection{Numerical sampling and discussion}
\begin{figure}[t]
	\includegraphics[width=0.5\textwidth]{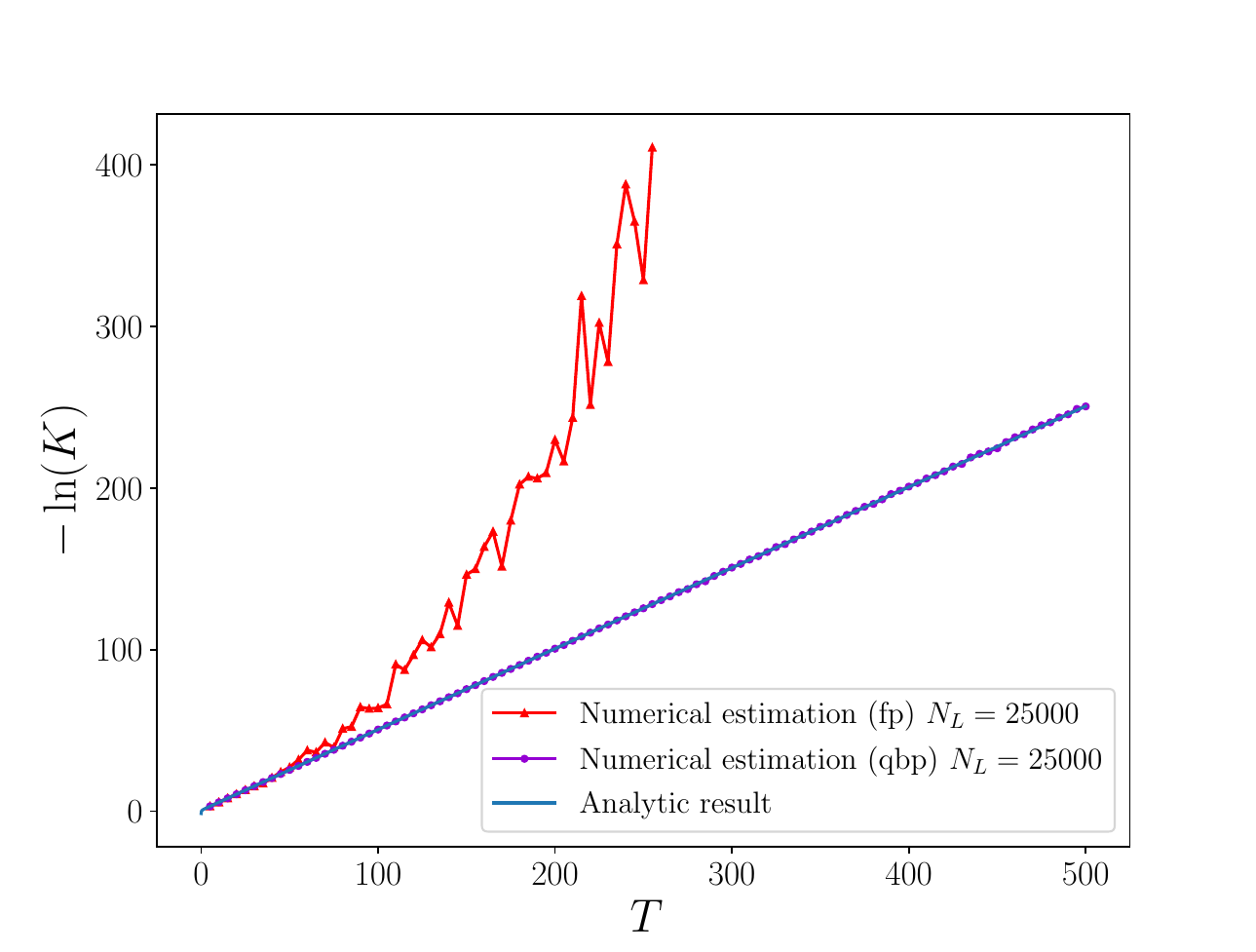}
	\caption{WMC estimate of the kernel $K(0,0 ; T)$ for the harmonic oscillator ($\omega = 1$, $N_L=25000$, $N_p=5000$, $\Omega=0.75$) using free particle (fp) paths and trajectories generated in a quadratic background potential (qbp), compensated via (\ref{eqSimOmegaF}).}
	\label{figHOHO}
\end{figure}

\begin{figure}[t]
	\includegraphics[width=0.5\textwidth]{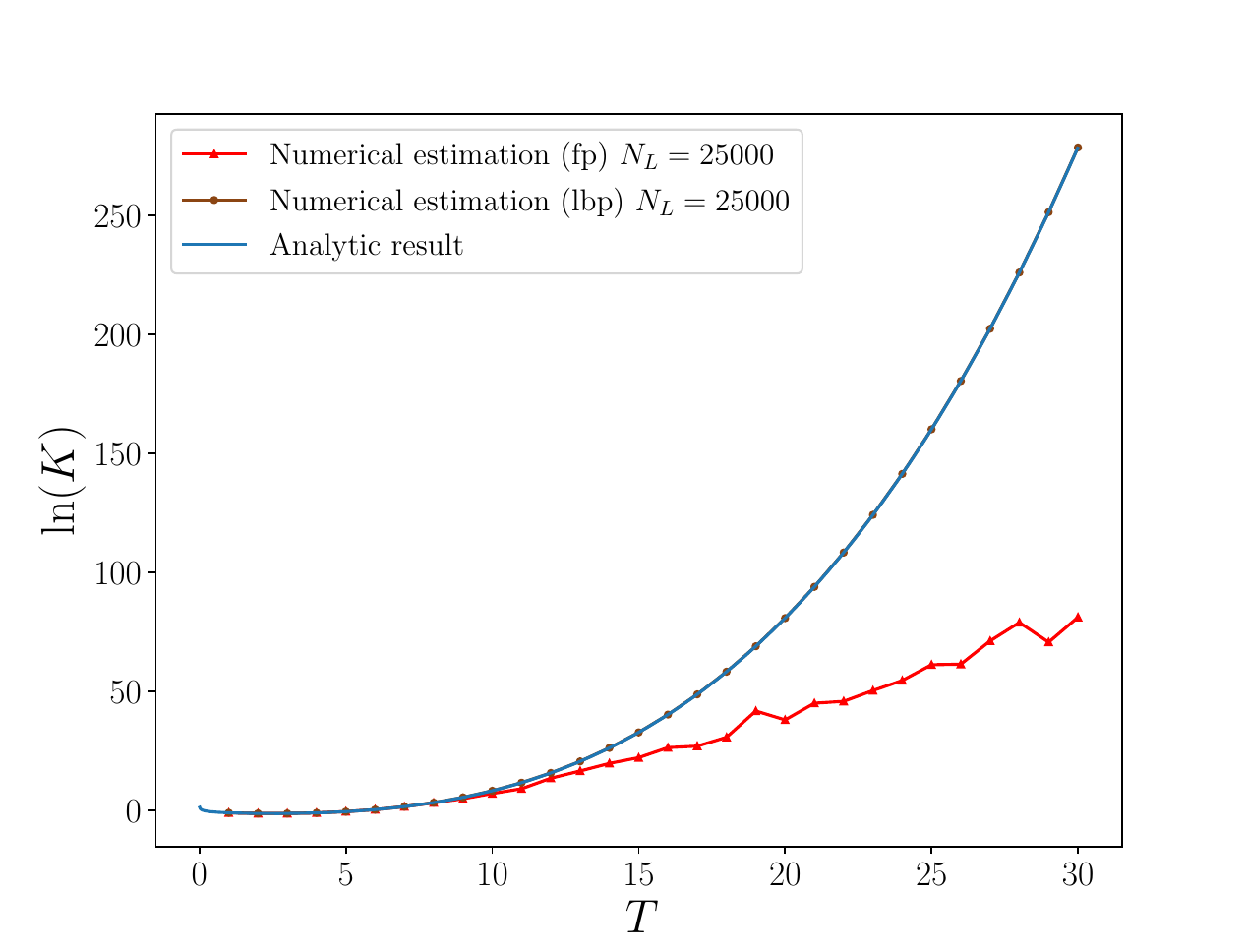}
	\caption{WMC estimate of the kernel $K(0,0 ; T)$ for the linear potential ($k = 0.5$, $N_L=25000$, $N_p=5000$, $\kappa=0.48$) using free particle (fp) trajectories and trajectories generated in a linear background potential(lbp), compensated via (\ref{eqSimkappaG}).}
	\label{figLinLin}
\end{figure}

In figures \ref{figHOHO} and \ref{figLinLin} we compare illustrative WMC estimates of the kernel with and without a background potential (henceforth we set $m=1$ throughout). The compensation factors ensure that the kernel is correctly reproduced for short times, agreeing both with previous estimations and the known analytic results. Note, however, that the simulations in background potentials extend the time interval in which the analytic result is reproduced by at least an order of magnitude. 

The numerical simulations also provide data on the distributions of values of the Wilson line variables in the backgrounds. They directly verify the relations (\ref{eqPvOmegaF}) and (\ref{eqPvkappaG}) for trajectories generated in the backgrounds, and of the ``shifted'' distributions (\ref{eqPvPrime}) and (\ref{eqPvPPrime}) after compensation as in (\ref{eqSimOmegaF}) and (\ref{eqSimkappaG}). The results of this analysis can be found in Appendix \ref{secApPAP}. They validate the algorithms in Appendix \ref{secApAlg} and the analytic determination of the adjusted PAPs given in (\ref{eqPvPrime}) and (\ref{eqPvPPrime}).

Analysis of these results reveals how undersampling is overcome. The background potentials influence the spatial distribution of trajectories to concentrates them about the regions where the Wilson lines, $\e^{-v[x]}$, provide significant contributions to estimation of the path integral. This is reflected in the modifications to the distribution, $\Pv(v)$, from trajectories generated by the new algorithms: smaller values of $v$ become more likely (see Appendix \ref{secApPAP}) which improves the sampling of the potential. The systematic bias incurred by modifying the distribution on the $v[x]$ is removed by the compensating factors in the Monte Carlo estimate.

The method has proven versatile to changes in the parameters of the systems under study (we discuss this in the next section). However, it is apparent that the compensating factor will only be obtainable for especially simple systems (see (\ref{eqPvOmegaF}-\ref{eqPvkappaG})). Even in backgrounds with the same functional form as the potential, say $U_{V}(x) \equiv \mu V(x)$ so that (\ref{eqPvOmegaF}-\ref{eqPvkappaG}) are replaced by the general result
\begin{equation}
	\Pv_{\mu V}(v | y, x; T) = \frac{K_{0}(y, x; T)}{K_{\mu V}(y, x; T)}\, \e^{-\mu v}\, \Pv(v | y, x; T)\,,
\end{equation}
it will be likely that (a) the PAP $\Pv(v | y, x; T)$ is unknown for this system and / or (b) it is non-trivial to generate trajectories in said background. Hence, a more universal approach was developed, outlined in the next section.

\section{Compensating potential}
\label{secPot}
We present a method that exploits our ability to generate trajectories in quadratic or linear backgrounds (see Appendix \ref{secApAlg}) for application to \textit{any} physical potential, $V(x)$. Here, the compensation for the background potentials is done numerically, evading the need to determine the PAP analytically as in the previous section.  

As outlined in Appendix \ref{secApAlg}, generating trajectories in the backgrounds $U_{\Omega}$ or $U_{\kappa}$ is formally equivalent to adding an unwanted term to the free particle action, leading to the incorrect estimations in (\ref{eqSimOmega}) and (\ref{eqSimkappa}). Clearly, the desired action can be restored by subtracting the unwanted term by hand:
\begin{align}
	\label{eqKvOmegaAn}
	K_{V} &= K_{\Omega}\Big\langle \e^{-\int_{0}^{T} d\tau\, \big[V(x(\tau)) - \frac{1}{2}m\Omega^{2}x(\tau)^{2} \big]} \Big\rangle_{\Omega}\\
	K_{V} &= K_{\kappa}\Big\langle \e^{-\int_{0}^{T} d\tau\, \big[ V(x(\tau)) - \kappa x(\tau) \big] } \Big\rangle_{\kappa}\,,
	\label{eqKvKappaAn}
\end{align}
where the new normalisation factors take into account the background potentials. Then, simulations will provide Monte Carlo estimations of the kernel via
\begin{align}
	\hspace{-0.9em}K_{V} &\MC K_{\Omega} \jsum{\nu_{i}^{\Omega} \sim \Pvt(v)}{i=1}{N_{L}} \e^{-\nu_{i}^{\Omega}} \MC K_{\Omega} \int_{-\infty}^{\infty}\, \Pvt(v)\,\e^{-v}dv
	\label{eqKvModOmega}
\end{align}
and the analogous relation
\begin{align}
	\hspace{-0.9em}K_{\kappa} &\MC K_{\kappa} \jsum{\nu_{i}^{\kappa} \sim \Pvh(v)}{i=1}{N_{L}} \e^{-\nu_{i}^{\kappa}} \MC K_{\kappa} \int_{-\infty}^{\infty}\, \Pvh(v)\,\e^{-v}dv\,.
	\label{eqKvModKappa}
\end{align}
It remains to determine the new PAP, taking into account the subtracted potential in the action.
\begin{figure}[b]
	\includegraphics[width=0.5\textwidth]{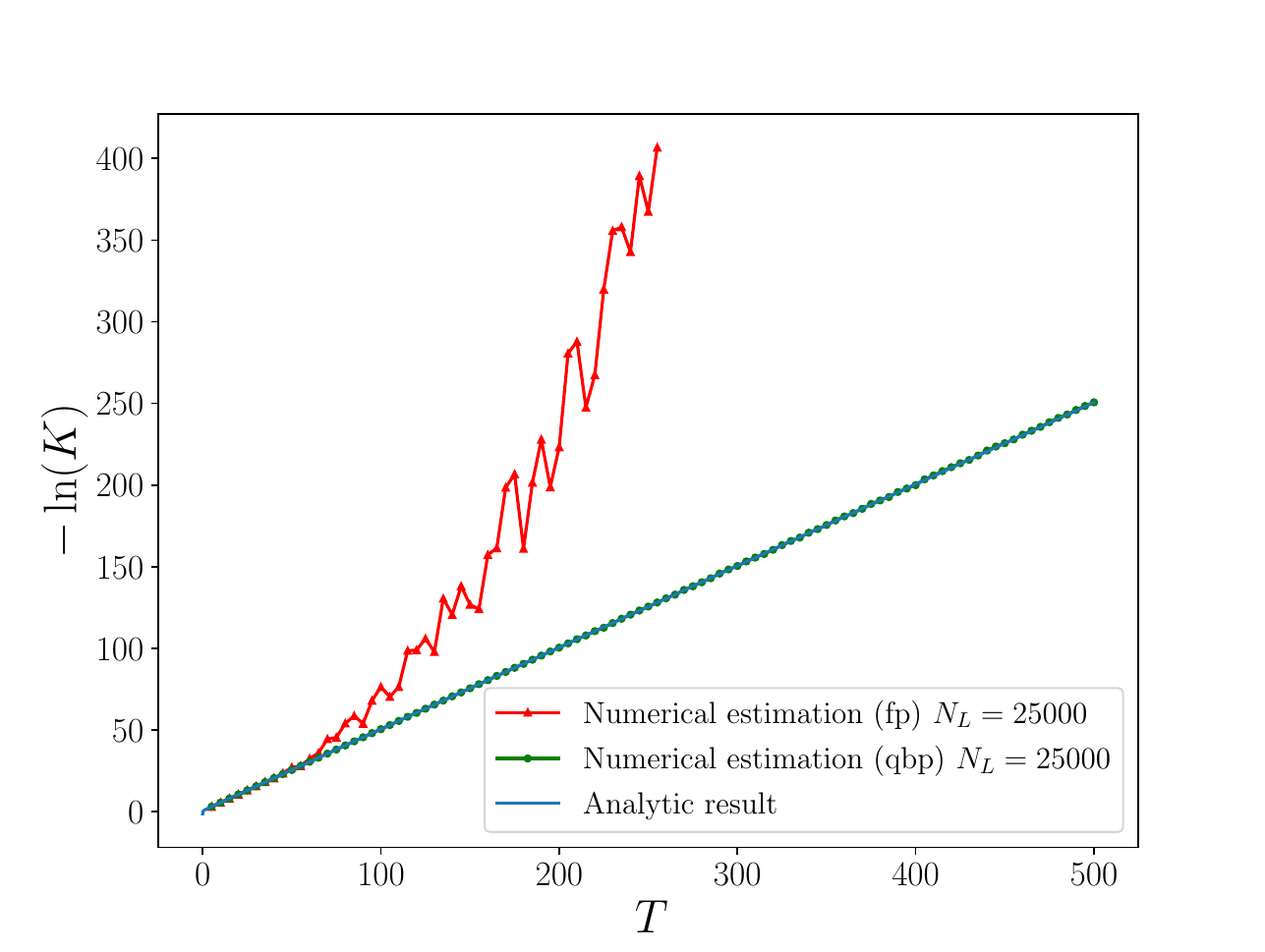}
	\caption{WMC estimate of $K(0,0 ; T)$ for the harmonic oscillator ($\omega = 1$, $N_L=25000$, $N_p=5000$, $\Omega=0.75$) using fp trajectories and a qbp, with the potential subtraction (\ref{eqKvOmegaAn}).}
	\label{figHOHOAction}
\end{figure}
\subsection{Wilson Line distributions}
The distribution on the $\nu^{\Omega}$ is given by constraining values of the line integral of the ``effective potential'' on trajectories in the harmonic oscillator background, $\Pvt(v) = \big\langle \delta \big( v - \nu[x] \big) \big\rangle_{\Omega}$, which we determined to be
\begin{align}
	\hspace{-2em}	\Pvt(v) &= \frac{\omega^{2}}{\omega^{2} - \Omega^{2}}\, \Pv_{\Omega}\big( \frac{\omega^{2}}{\omega^{2} - \Omega^{2}}\, v\big) \\
	\hspace{-2em} &= \sqrt{\frac{\sinh(\Omega t)}{\Omega t}}  
	\frac{\omega^{2}}{\omega^{2} - \Omega^{2}}  \e^{-\frac{\Omega^{2}}{\omega^{2} - \Omega^{2}}v} \Pv\Big(\frac{\omega^{2}}{\omega^{2} - \Omega^{2}} \,v\Big)\,.
	\label{eqPvt}
\end{align}
With this we can show that (\ref{eqKvModOmega}) indeed gives a correct Monte Carlo estimation by changing the integration variable to $v' = \frac{\omega^{2}}{\omega^{2} - \Omega^{2}} \,v$, yielding
\begin{equation}
	\hspace{-0.5em}K_{\Omega} \int_{-\infty}^{\infty}\, \Pvt(v)\,\e^{-v}dv = K_{\Omega} \frac{K_{0}}{K_{\Omega}} \int_{-\infty}^{\infty}\, \Pv(v^{\prime})\, \e^{-v^{\prime}}dv^{\prime} = K_{V}\,,
\end{equation}
as desired. 

Similarly, we calculate how the $\nu^{\kappa}$ are distributed through the constrained path integral taking the linear background into account: $\Pvh(v) = \big\langle \delta \big( v - \nu[x] \big) \big\rangle_{\kappa}$, giving
\begin{align}
	\hspace{-1em}\Pvh(v) &= \frac{k}{k-\kappa}\, \Pv_{\kappa}\big(\frac{k}{k-\kappa}\,v\big)	 \\
	\hspace{-1em} &=  \e^{\frac{\kappa T}{2}(x+y) - \frac{\kappa^{2}T^{3}}{24}}\, 
	\frac{k}{k - \kappa} \, \e^{-\frac{\kappa}{k - \kappa}v} \,\Pv\Big(\frac{k}{k - \kappa} \,v\Big)\,.
	\label{eqPvh}
\end{align}
After changing variables to $v' = \frac{k}{k - \kappa}v$, (\ref{eqPvh}) proves that (\ref{eqKvModKappa}) is a faithful Monte Carlo estimation of the kernel.

WMC simulations were carried out to confirm the predictions of the above analysis. In particular, the modified PAPs in (\ref{eqPvt}) and (\ref{eqPvh}) were sampled by probing the values of the Wilson line variable for trajectories produced in the backgrounds. We present results of this sampling in Appendix \ref{secApPAP}, which confirm the theoretical determination presented above. 
\begin{figure}[b]
	\includegraphics[width=0.5\textwidth]{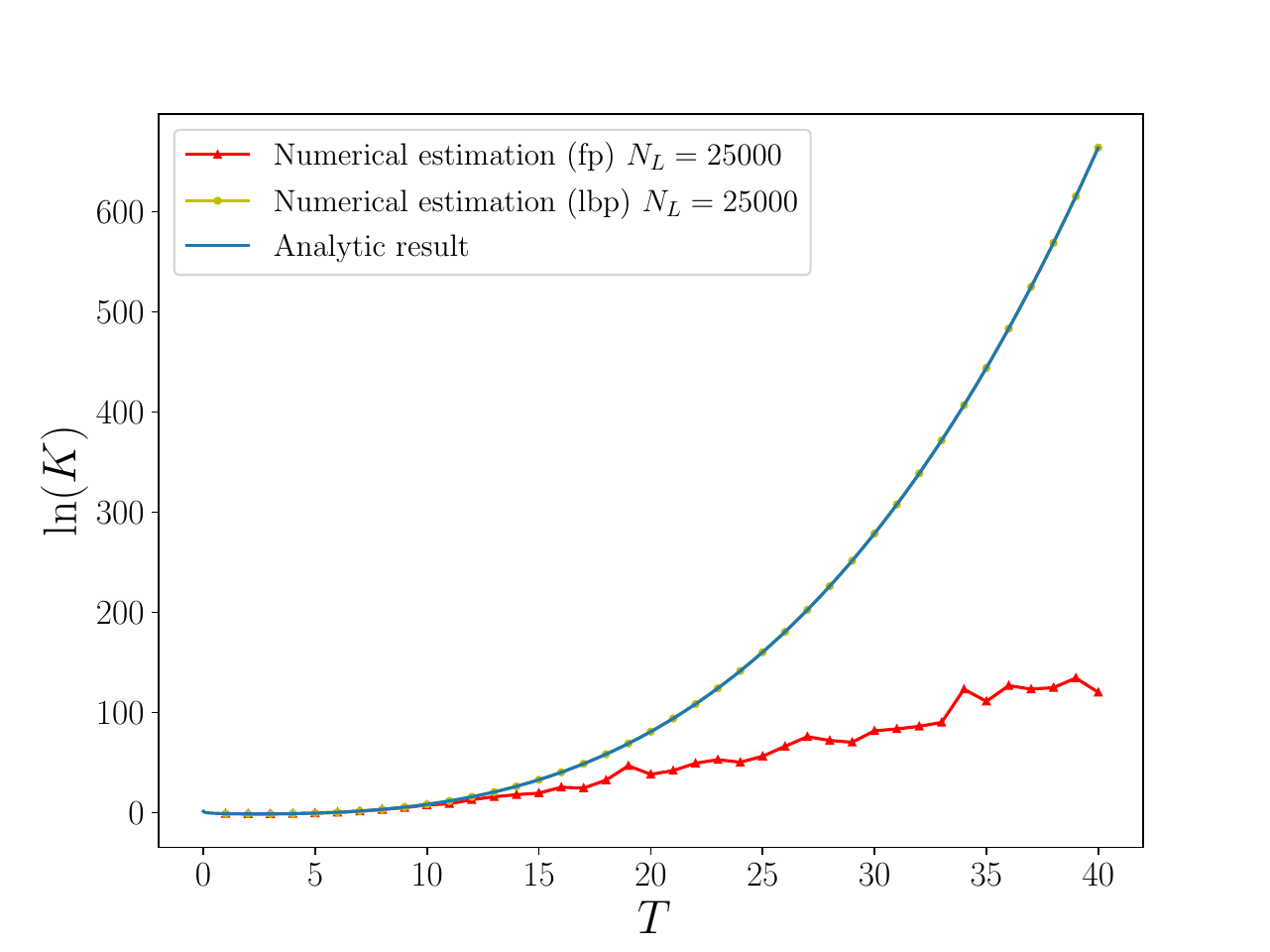}
	\caption{WMC estimate of $K(0,0 ; T)$ for the linear potential ($k = 0.5$, $N_L=25000$, $N_p=5000$, $\kappa=0.48$) with fp trajectories and a lbp, compensated with potential subtraction (\ref{eqKvKappaAn}).}
	\label{figLinLinAction}
\end{figure}
Here we instead focus on showing the improvement in the estimation of the propagator for the quantum systems, overcoming the undersampling problem that limited previous work. As a first step, we estimated the kernel for the same potentials (i.e. harmonic oscillator and linear potential) as for the analytic approach of the previous subsection. This allows us to confirm that the subtraction in the action gives a comparable improvement in precision that extends the range of times for which the simulations provide good estimates of the kernel -- see figures \ref{figHOHOAction} and \ref{figLinLinAction}.

A natural question is whether the undersampling problem reappears, albeit for larger transition times. Indeed, for a fixed value of the background parameters ($\Omega$ and $\kappa$), a deviation from linearity can eventually occur. It is still associated to the growth in the spatial extent of the trajectories driven by the Gaussian distribution on velocities overcoming the confining effect of the background potentials. But it can easily be mitigated by simply increasing the value of $\Omega$ or $\kappa$ as appropriate for the system under study. Doing so strengthens the background potential and ensures that it continues to compensate for the diffusion of trajectories. This takes advantage of the fact that the error induced by discretising the Riemann integral of the compensating potential is much smaller than that caused by discretising the path integral over trajectories. 
\begin{figure}[b]
	\includegraphics[width=0.5\textwidth]{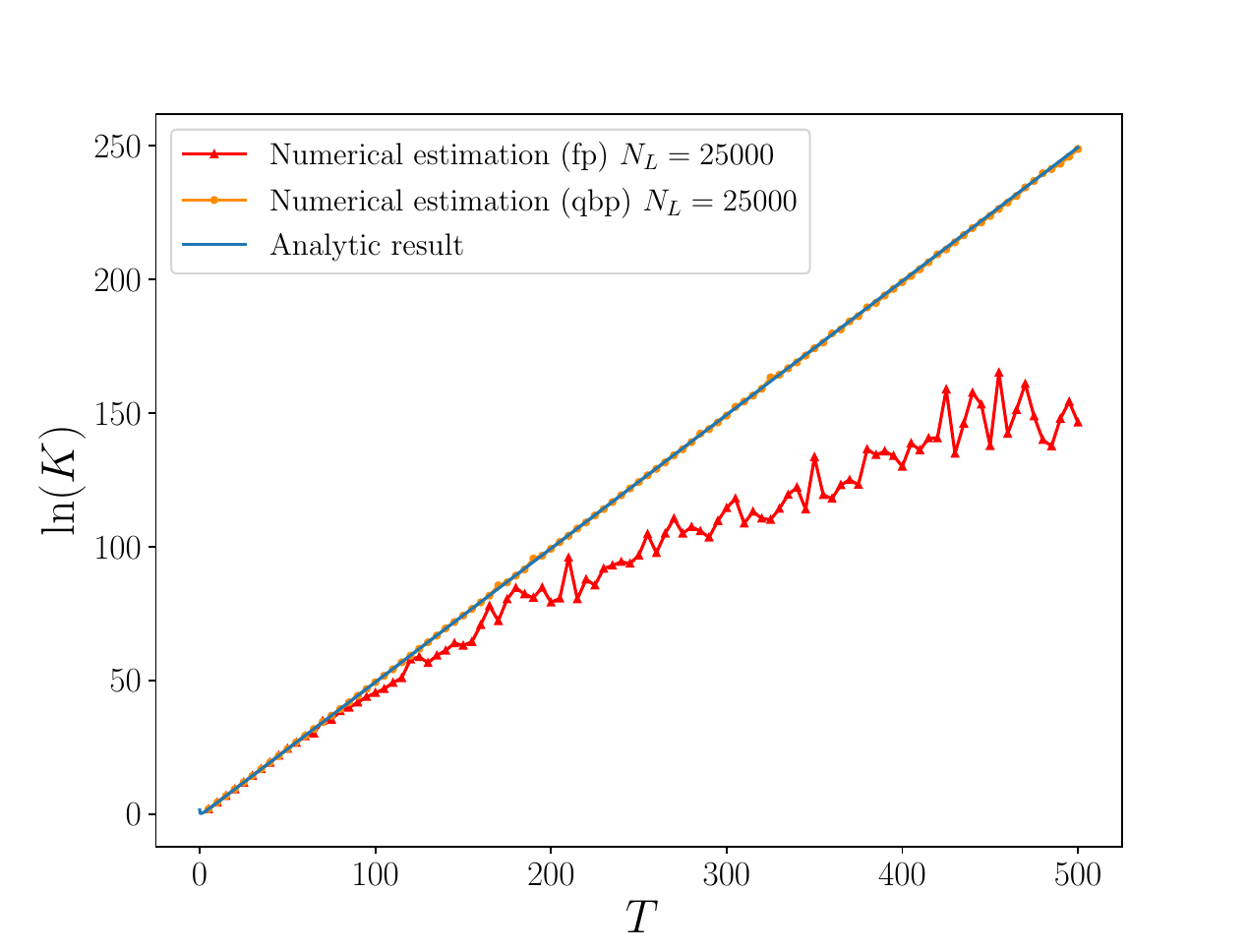}
	\caption{WMC estimate of the kernel $K(0,0 ; T)$ for the Pöschl-Teller potential ($\lambda = 1$, $\alpha = 1$, $N_L=25000$, $N_p=5000$, $\Omega=0.75$) using fp trajectories and qbp with potential subtraction.}
	\label{figPT}
\end{figure}
\section{Applications}
\label{secApps}
So far, we have focussed on confirming the theoretical advances of earlier sections, using especially simple systems and related, analytically convenient background potentials. In this section we analyse some less trivial quantum systems (where sufficient analytic knowledge still exists) to test the proposed method: the \textit{Pöschl-Teller} ``reflectionless potential,'' the absolute value potential and a system with a cubic memory kernel. In the symmetric cases, WMC trajectories are generated in a harmonic background centered about the minimum of the potential, and the potential subtraction scheme is applied; for the cubic memory kernel system, we generate the trajectories in a linear background centered at the origin in order to favour trajectories exploring the negative real axis.

We briefly comment on errors in the proceeding simulations. As outlined by \cite{UsMonteCarlo, Gies:2001zp, Gies:2001tj}, for $N_{L}$ sufficiently larger than $N_{P}$ the statistical error from the path integral discretisation dominates the systematic error in calculating the (Riemann) integral of the potential along the trajectories. Then the error in the estimation of the kernel is well approximated by the standard error in the mean of values of the Wilson line. The percentage error in the simulations below was found to be of order $2$-$4\%$ (unless otherwise stated) -- such small error bars have been suppressed for clarity. Autocorrelation is avoided by using an independent set of trajectories for each value $T$. 

\subsection{Estimating the propagator and ground state energies}
The propagator for the modified Pöschl-Teller potential, $V_{\lambda}(x) = -\frac{\alpha^{2}}{2m} \frac{\lambda(\lambda + 1)}{\cosh^{2}(\alpha x)}$,  is known in closed form \cite{Grosche}. We simulate this for coupling $\alpha = 1$ and with $\lambda = 1$ bound state (making it most susceptible to the undersampling problem), using a harmonic oscillator background with centre $x = 0$. Since the PAP is unknown for this system, a compensating factor cannot be determined analytically, so we rely on the potential subtraction scheme. Figure \ref{figPT} compares our results using the algorithms reported here favourably to those in \cite{UsMonteCarlo}.

\begin{figure}[b]
	\includegraphics[width=0.5\textwidth]{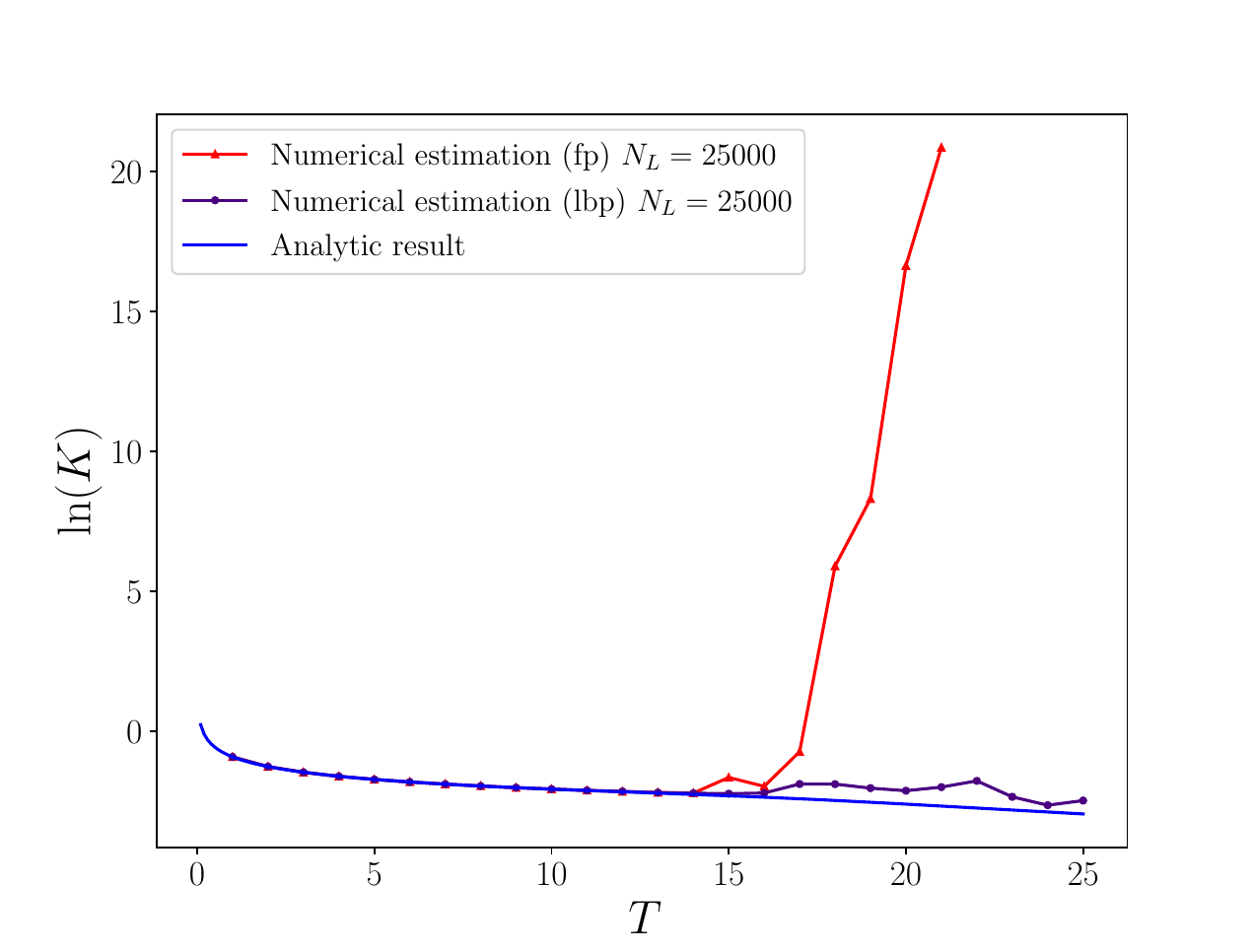}
	\caption{WMC estimate of the kernel $K(0,0 ; T)$ for a cubic memory kernel ($\mu = 0.025$, $N_L=25000$, $N_p=5000$, $\kappa=0.1$) using fp and lbp trajectories (potential subtraction).}
	\label{figRM}
\end{figure}

The background potential is seen to overcome the undersampling problem, allowing simulations up to transition times at least 8 times greater. Then, using (\ref{eqKlimit}), we estimate the ground state energy through a linear fit to the propagator for $t \in [5, 50]$. We find $E_{0} = -0.50004391$ compared to the analytic result $E_{0} = -0.5$, which represents an order of magnitude increase in precision over previous WMC estimations.

We treat the the absolute value potential, defined by $V_{\kappa}(x) = \kappa|x|$, in Appendix \ref{secApA}, reporting simply the final estimation of the ground state energy for parameters $m = 1$, $\kappa = 0.5$ here. We find an estimation $E_{0} = 0.50939336$ for a linear fit to the propagator in the range $T \in [5, 30]$ (analytic result: $E_{0} = 0.509397\ldots$).

Finally, to demonstrate the use of a linear background, we consider a system with a cubic memory kernel \cite{Grosche}, such that $S[x] = \int_{0}^{T}d\tau \big[m\frac{\dot{x}^{2}}{2} - \mu \big( \int_{0}^{T}d\tau\, x(\tau)\, \big)^{3} \big]$ with $\mu$ constant. This provides a linear asymmetry but with a cubic nonlinearity. The propagator can be found in closed form as (we take $x = y = 0$ for convenience)
\begin{equation}
	\hspace{-1em}K(0, 0; T) = \frac{2}{T^{2}} \Big( \frac{\sqrt{3}}{\mu} \Big)^{\frac{1}{3}}  \mathrm{Ai}\Big[ \frac{4}{T^{6}} \Big(\frac{\sqrt{3}}{\mu}\Big)^{\frac{4}{3}} \Big]\, \e^{\frac{16}{T^{9}\mu^{2}}},
\end{equation}
where $\textrm{Ai}$ is the Airy function. For $\mu > 0$ the path integral is strongly weighted by trajectories in the region $x > 0$, where the potential is largest. As for the linear potential, however, Monte Carlo estimation of the kernel suffers severe undersampling due to the sum in (\ref{eqKMC}) being dominated by a small number of trajectories for which  $\int_{0}^{T}d\tau\, x(\tau)$ is excessively large. This can be seen in the simulations using free particles (fp) in Figure \ref{figRM}. In this case, then, we use a linear background to encourage trajectories towards more \textit{negative} values of $x$. 

Although Figure \ref{figRM} clearly shows that the estimation of the propagator is substantially improved, especially for smaller values of $T$, in this case we found greater instability in the predictions obtained with the background potential method as $T$ increases, due to favouring trajectories in a lower-importance region in order to avoid undersampling. Indeed, here for $T < 15$ the percentage error is $\mathcal{O}(10\%)$, but varies from $5\%$ to $20\%$ for larger values of T. As demonstrated in \cite{UsMonteCarlo}, this instability can be partially overcome by (a) increasing $N_{L}$ and (b) averaging over a suitable number of repeated simulations. It can also be reduced by adaptively varying the value of $\kappa$ as the transition time increases. The instabilities found for this system are intended to be clarified in future work.


\section{Conclusion}
In this article we have presented two methods for improving Monte Carlo simulations of the quantum mechanical propagator, along with accompanying algorithms (in Appendix A) that generate point particle trajectories in appropriate background potentials. We have confirmed the correctness of these Monte Carlo estimates, both analytically and numerically, and shown how they overcome an undersampling problem that has previously hindered simulation of the Schrödinger kernel for large transition times, thereby limiting the precision of estimations of physical quantities such as energy levels. 

The methods reported here allow us to extend the range of transition times accessible to simulations by an order of magnitude, consequently improving estimations of the ground state energies of the systems under study, again by an order of magnitude. We expect that these methods will be widely applicable to more general systems, including singular potentials \cite{UsMonteCarlo}, for (spatially dependent) electromagnetic fields \cite{Gies:2001zp} or in curved space \cite{Corradini:2020tgk}. 

Note that this work also clarifies why the ``regulating mass'' in \cite{Corradini:2020tgk} had a minimal effect on their numerical results. There, a small $\alpha$ (see (\ref{eqAlphaOmega})) was used at short times, corresponding to $\Omega T \ll 1$ and $\frac{\Omega^{2}}{\omega^{2}}v \ll 1$. For these parameters, $F(v) \approx 1$ so that $\Pv_{\Omega}(v) \to \Pv(\Omega)$ and the Monte Carlo estimation remains approximately faithful. We have verified that even the free particle propagator can be simulated for the first time with the algorithms proposed here, by using potential subtraction in either background for an otherwise non-interacting system.

The extension of this work to relativistic particle trajectories relates, via the worldline formalism, to studying the propagator (open worldlines) or effective action (closed trajectories) for quantum fields. In this context, the formation of bound states can be examined by considering a multi-particle interacting system. Such inter-particle interactions are often highly localised, or even singular, where the importance sampling presented here would be of considerable benefit for accelerating the convergence of numerical simulations. Of course the inclusion of spin degrees of freedom is an important aspect of numerical simulations. In the case of a magnetic moment coupling to a magnetic field, for instance, we would aim to incorporate information about the spatial variation of the magnetic field to achieve a similar importance sampling of this interaction, as demonstrated here for spatially localised potentials.

\bigskip

\section*{Acknowledgements}
The authors thank CONACyT for financial support; IA though a scholarship and JPE via ``Ciencia de Fronteras'' project \#\!\! 170724. They are indebted to Christian Schubert, Idrish Huet \& Anabel Trejo for helpful discussions, and thank Olindo Corradini for useful suggestions. They also thank an anonymous referee for helpful comments regarding statistical uncertainties. 

\onecolumngrid
\appendix

\pagebreak
\newpage
\section{Monte Carlo algorithms}
\label{secApAlg}
In this appendix we present the numerical algorithms for generating trajectories in quadratic and linear backgrounds. The method is based on discretising (\ref{eqPqU}) for $U_{\Omega}$ and $U_{\kappa}$ and diagonalising the result.
\subsection{Harmonic oscillator background}
An algorithm was introduced in \cite{Corradini:2020tgk} to generate trajectories with a regularising ``mass term'' (quadratic in the field $x$). Denoting by $Y$ the sum in the exponent of (\ref{eqPq}),  \cite{Corradini:2020tgk} considered the modification:
\begin{equation}
	Y\,\rightarrow\,Y^{(\alpha)}=\sum_{k=1}^{N_p}[(q_k-q_{k-1})^2+\alpha q_k^2]\,,\quad\alpha>0\,.
\end{equation}
This can be (non-orthogonally) diagonalised by:
\begin{eqnarray}
	Y^{(\alpha)}&=&\sum_{k=1}^{N_p-1}C_{N_p-k}^{(\alpha)}\bar{q}_k^2\,,
\end{eqnarray}
with the identification
\begin{equation}
	\bar{q}_k=q_k-\frac{1}{C_{N_p-k}^{(\alpha)}}q_{k-1}\, ,\; k=1,2,\ldots,N_p-1\,,
\end{equation}
and with $C_{k}^{(\alpha)}=C_1^{(\alpha)}-\frac{1}{C_{k-1}^{(\alpha)}}$\, ($C_1^{(\alpha)}=2+\alpha$). In this work we instead interpret the $\alpha q_{k}^{2}$ term in $Y^{(\alpha)}$ as providing a genuine background potential, $U_{\Omega} = \frac{1}{2}m\Omega^{2}x^{2}$, that concentrates trajectories about its minimum. The frequency of this harmonic potential is related to the mass parameter in the continuum limit by
\begin{equation}
	\alpha=\frac{\Omega^2 T^2}{N_p^2}\,. \label{eqAlphaOmega}
\end{equation}
Then the corresponding algorithm reads as follows:
\begin{enumerate}
	\item Generate $N_p-2$ vectors $\omega_i$, $i=1,2,\ldots,N_p-1$, distributed according to $\mathcal{P}(\omega_i)\propto \exp(-\omega_i^2)$.
	\item Compute unit vectors $\bar{q}_i=\sqrt{\frac{2}{N_p C_{N_p-i}^{(\alpha)}}}\,\omega_{i-1},$ for \newline $ i=1,2,\ldots,N_p-1.$
	\item Construct the unit loop according to 
	\begin{eqnarray}
		\nonumber q_1&=&\bar{q}_1,\\
		q_i&=&\bar{q}_i+\frac{1}{C_{N_p-i}^{(\alpha)}}\,q_{i-1},\,\; i=2,3,\ldots,N_p-1\, .
	\end{eqnarray}
	\item Repeat the process $N_L$ times. 
\end{enumerate}
We describe in the main text how to compensate for the unwanted bias caused by this modification to	 trajectories. 


\subsection{Linear background}
For one-side dominated potentials, it is instead favourable to produce trajectories in a linear background potential, so we let
\begin{equation}
	Y\,\rightarrow\,Y^{(\beta)}=\sum_{k=1}^{N_p}[(q_k-q_{k-1})^2+2\beta q_k]\,,\quad\beta>0\,.
\end{equation}
This time the diagonalisation is achieved by
\begin{equation}
	Y^{(\beta)} = \sum_{k=1}^{N_p-1}C_{N_p-k}^{(\beta)}\,\bar{q}_k^2-\sum_{k=1}^{N_p-1}\frac{\beta_{k}^{2}}{C_k^{(\beta)}}\,. \label{LPB1}
\end{equation}
where
\begin{eqnarray}
	\nonumber \bar{q}_k&=&q_{k}-\frac{1}{C_{N_p-k}^{(\beta)}}\big(q_{k-1}-\beta_{N_p-k}\big),\;k=1,2,\ldots,N_p-1\,, 
\end{eqnarray}
with $C_{k}^{(\beta)}=\frac{k+1}{k}$ and $	\beta_{k}=\frac{k+1}{2}\beta$.
The final sum in \eqref{LPB1} can be evaluated to give 
\begin{equation}
	\sum_{k=1}^{N_p-1}\frac{\beta_{k}}{C_k^{(\beta)}}=\frac{\beta^2}{12}N_p(N_p+1)(N_p-1)\,. \label{LPB2}
\end{equation}
In the continuum limit, the linear term in $Y^{(\beta)}$ corresponds to a potential $U_{\kappa}(x) = \kappa x$ where 
\begin{equation}
	\beta = \frac{\kappa}{m}\frac{T^{2}}{N_{P}^{2}}\,, \label{eqBetaKappa}
\end{equation}
and the sum in equation (\ref{LPB2}) tends to $\frac{\kappa^{2}T^{3}}{24 m}$. The numerical algorithm for this background follows the previous one with $C^{(\alpha)}_{k} \rightarrow C^{(\beta)}_{k}$, except that step 3 becomes:
\begin{enumerate}
	\item[3.] Construct the unit loop according to 
	\begin{eqnarray}
		\nonumber q_1&=&\bar{q}_1,\\
		\nonumber q_i&=&\bar{q}_i+\frac{1}{C_{N_p-i}^{(\beta)}}\left(q_{i-1}-\sqrt{\frac{m}{t}}\beta_{N_p-i}\right),\; i=2,3,\ldots,N_p-1\,.\\
	\end{eqnarray}
\end{enumerate}
Again, the bias induced by the presence of the background is removed by the procedures described in the main text.

\section{Path Averaged Potentials}
\label{secApPAP}
Here we report the functional forms of the path averaged potential first studied in \cite{PvHz}. For the linear potential, $V(x) = k x$, the PAP can be determined analytically and is Gaussian, 
\begin{equation}
	\Pv(v | y,x;T) = \sqrt{\frac{6}{\pi k^2 T^3}}\e^{-\frac{3}{2T}(x+y)^{2}}\e^{-\frac{6}{k^{2}T^{3}} \big(v^2 - kT(x+y)v\big)} = \sqrt{\frac{6}{\pi k^2 T^3}}\e^{-\frac{6}{k^{2}T^{3}} \big(v - \frac{kT}{2}(x+y)\big)^{2}}\,.
	\label{eqSupPvLin}
\end{equation}
Note that $\int_{\infty}^{\infty} dv\, \Pv(v | y,x;T) = 1$ and that $\Pv(v | y,x;T)$ satisfies the boundary conditions $\lim_{T \rightarrow 0}\Pv(v | y,x;T) = \lim_{k\rightarrow 0}\Pv(v | y,x;T) = \delta(v)$, which is the free particle limit.

A spectral representation of the PAP for the harmonic oscillator, $V(x) = \frac{1}{2}m\omega^{2}x^{2}$,  is given (for $x = y = 0$, sufficient for this work) in their equation (33) which we can simplify a small amount to
\begin{equation}
	\Pv(v | y,x;T) = 32\Theta(v)\sqrt{\frac{\omega T}{ \pi^{2}}} \sum_{n \textrm{ even}} \frac{n! \, v_{n}^{\frac{3}{2}} \,\e^{-v_{n}} }{2^{n+\frac{1}{2}} (\frac{n}{2})!^{2} \left((n+\frac{1}{2}) \omega T \right)^{\frac{5}{2}}} \Big[\Big(v_{n} - \frac{3}{4} \Big)K_{\frac{1}{4}}\left(v_{n}\right) + v_{n} K_{\frac{5}{4}}\left(v_{n}\right)\Big]\,,
	\label{eqSupPvHO}
\end{equation}
where $v_{n} = \frac{\big[\big(n + \frac{1}{2}\big)\omega T\big]^{2}}{8v}$ and the $K_{\alpha}$ are the modified Bessel functions of the second kind. The series converges very rapidly and can be truncated at $n$ as small as $10$ without sacrificing precision. This function satisfies the same normalisation and boundary condition in the limit of vanishing coupling, $\omega$. Both functions can therefore be interpreted as probability distributions for the Wilson line variable, $v$, on the space of Brownian motion trajectories. They measure the relative contribution to the path integral of trajectories whose Wilson line variable evaluates to a the given value, $v$. The kernel is obtained by integrating this over $v$ with Gaussian weight (see Main Text).

As described in the Main Text, a good sampling of the path integral equates to generating trajectories such that the values of their associated Wilson lines sample well the $\Pv$ appropriate to the system. Much of the work in the main text is completed in this ``$v$-space'' which provides insight in a similar way to how Fourier (momentum) space can complement working in position space. 

\subsection{Numerical analysis of trajectories and distributions}
The distributions (\ref{eqSupPvLin}) and (\ref{eqSupPvHO}) were corroborated numerically in \cite{PvHz, UsMonteCarlo}. Here we provide analogous numerical confirmation of the ``shifted distributions'' denoted by $\Pv_{\Omega}(v)$ and $\Pv_{\kappa}(v)$ in section III of the Main Text -- see equations (22-23). These also allow us to obtain the distributions $\Pvp(v)$ and $\Pvpp(v)$ satisfied by the shifted variables, $\{v_{i}^{\prime}\}$ and  $\{v_{i}^{\prime\prime}\}$ in the Main Text -- equations (26) and (30). The shifted distributions can be found by using the Fourier representation of the $\delta$-function or by noting that the constraint ensures that 
\begin{equation}
	\frac{1}{K_{\bullet}} \int \mathscr{D}x(\tau) \,\delta\big(v - v[x]\big) \, \e^{-\int_{0}^{T}d\tau\big[ \frac{m\dot{x}^{2}}{2} +U_{\bullet}(x)\big]} = \frac{K_{0}}{K_{\bullet}} \e^{- (U_{\bullet} / V)v }\,.
\end{equation}

\begin{figure}
	\centering
	\includegraphics[width=0.329\textwidth, trim={1.1cm 0.35cm 1.5cm 1.1cm},clip]{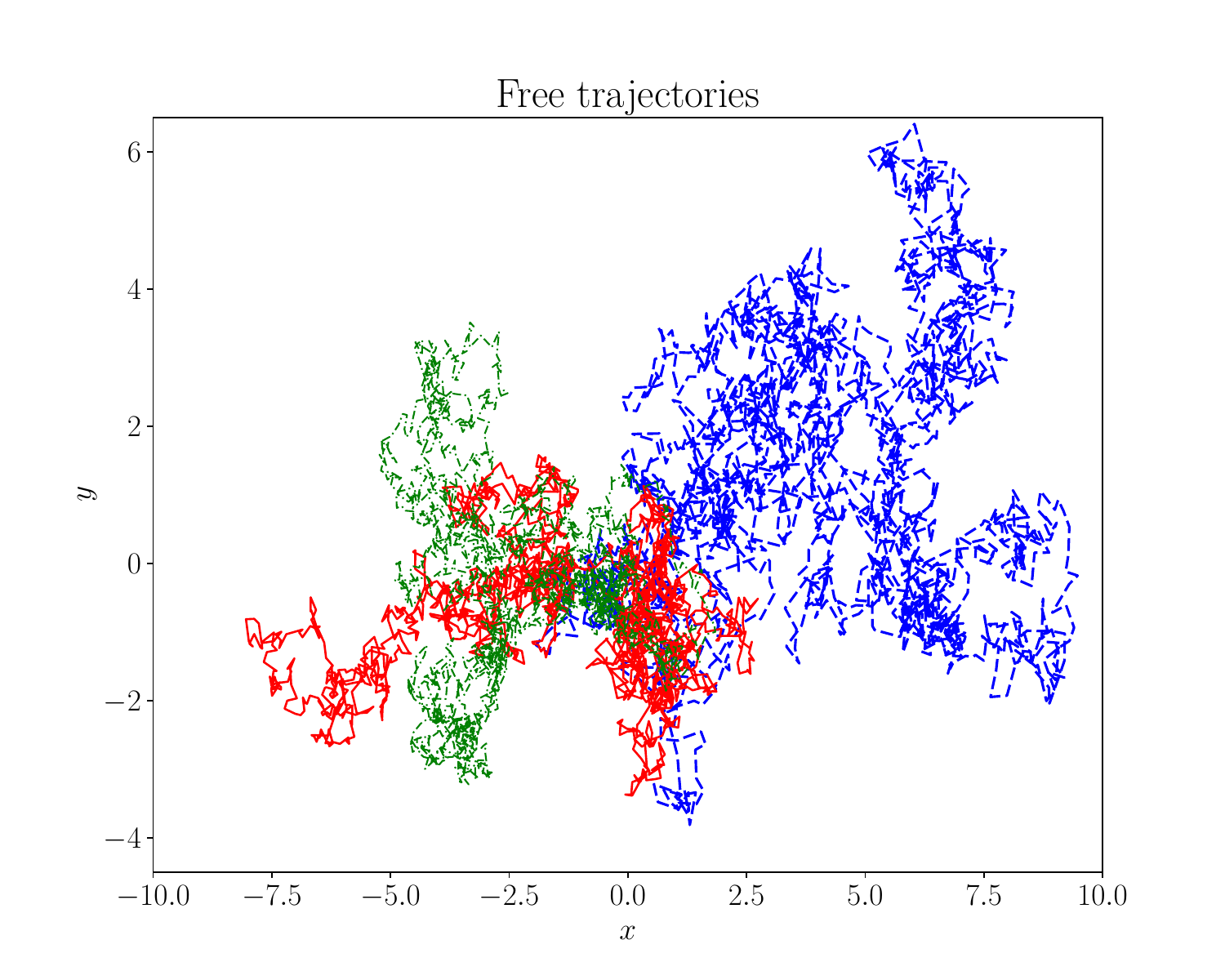}
	\includegraphics[width=0.33\textwidth, trim={1.1cm 0.35cm 1.5cm 1.1cm},clip]{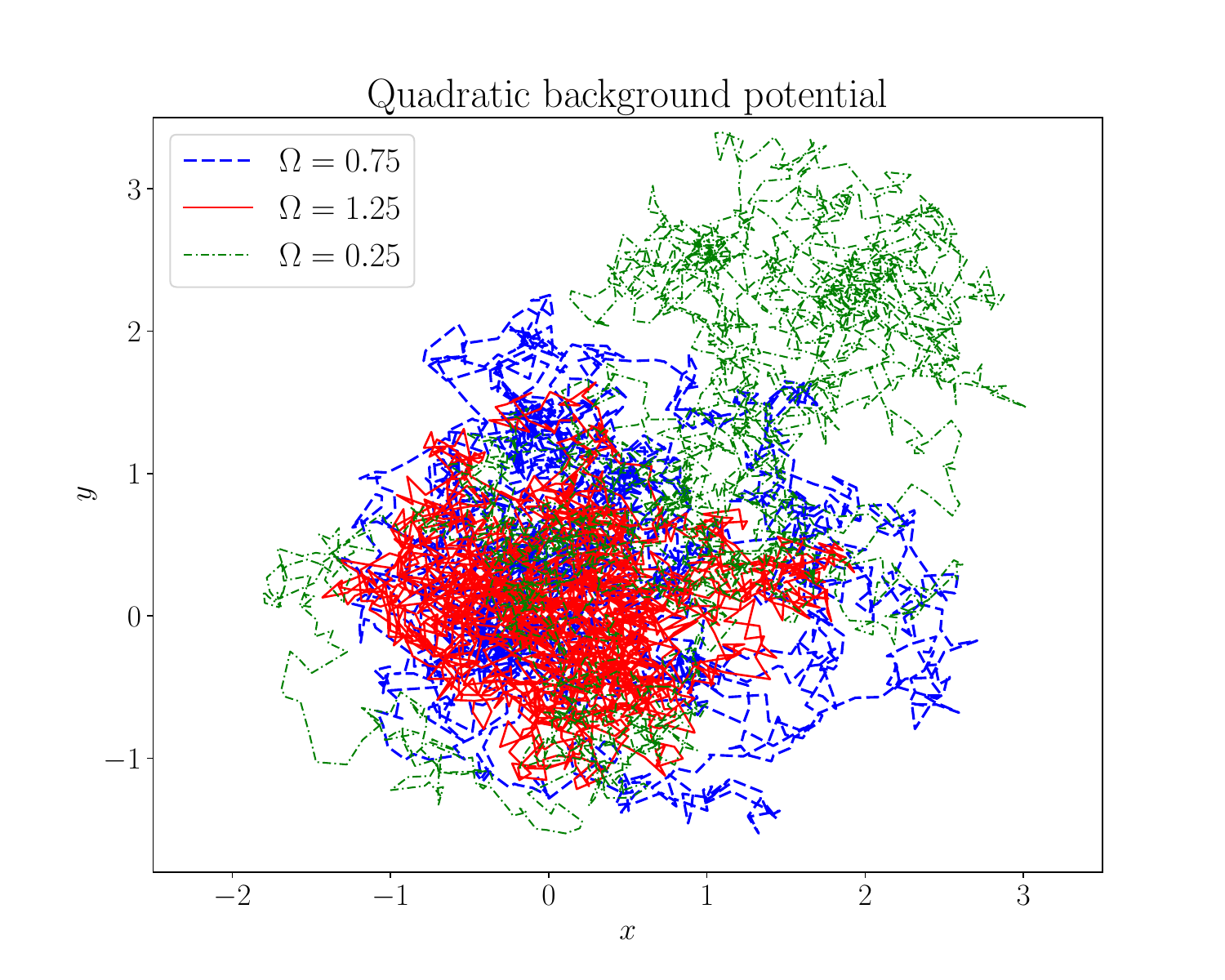}
	\includegraphics[width=0.33\textwidth, trim={1.1cm 0.35cm 1.5cm 1.1cm},clip]{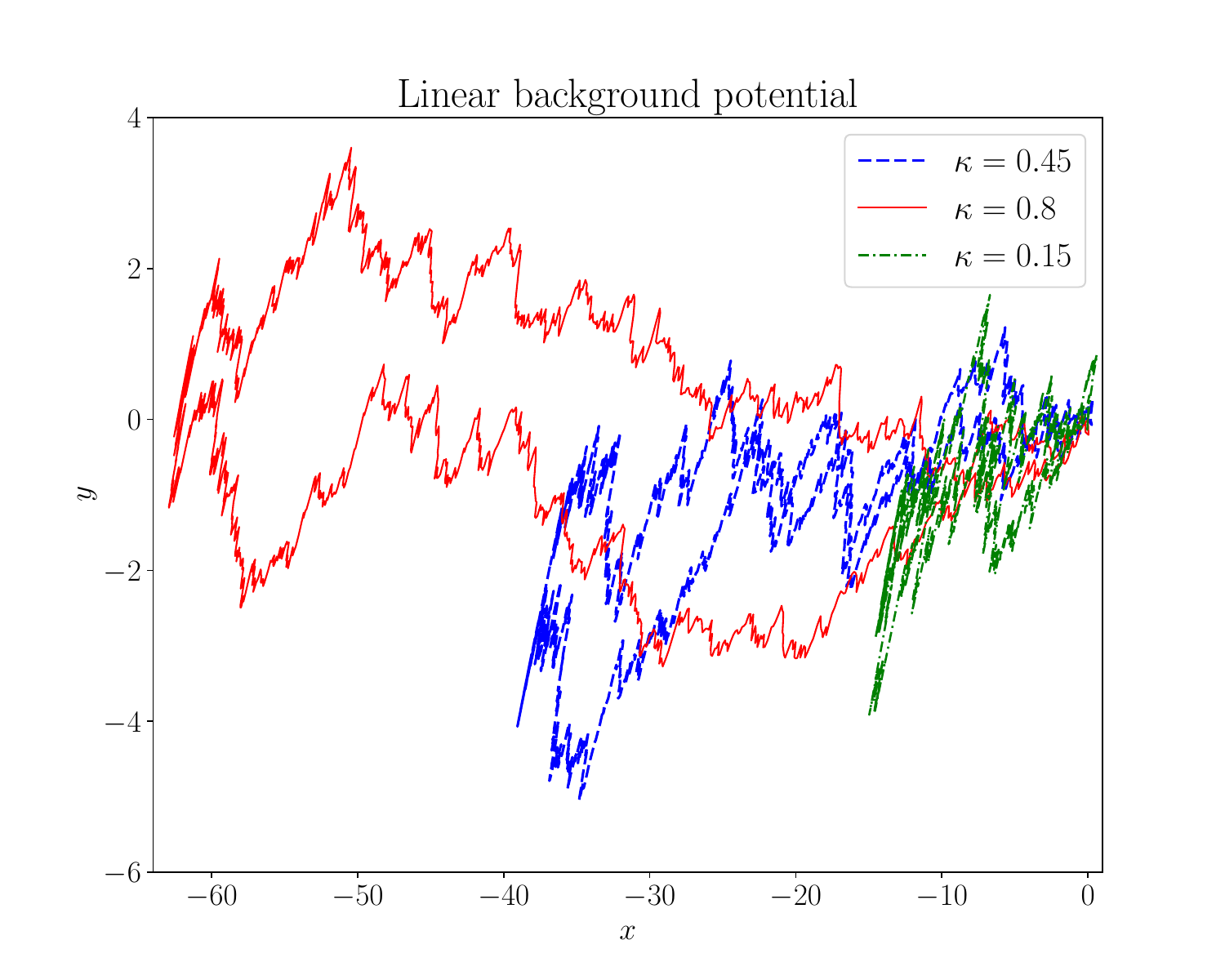}
	\caption{Comparison of trajectories generated using standard WMC (left panel) and including background potentials. Middle panel:  $U_{\Omega}(x) = \frac{1}{2}m\Omega^{2} x^{2}$ for $\Omega = 0.25$ (upper-most dashdotted green line), $\Omega = 0.75$ (central dashed blue line) and $\Omega = 1.25$ (left-most solid red line), $m = 1$ and $T = 20$. Right panel: for $U_{\kappa}(x) = \kappa x$ we used $\kappa = 0.15$ (right-most dashdotted green line), $\kappa = 0.45$ (central dashed blue line) and $\kappa = 0.8$ (far left solid red line), $m=1$ and $T = 25$. Free trajectories were generated with $m=1$ at $T=30$. All simulations used $N_{p} =2048 $ points per loop for each plot. The influence of the backgrounds on the spatial extent of trajectories is clear (note the differing scales on the axes).}
	\label{figLoops}
\end{figure}

We begin by showing how the spatial distribution of trajectories changes when the background potentials are present, using simulations in two dimensional quantum mechanics for illustrative convenience. The harmonic oscillator background should concentrate trajectories about its minimum, here taken to be the origin. The linear background (with respect to one dimension, $x$ say) will be seen to push trajectories towards smaller values of $x$. In Figure \ref{figLoops} we contrast trajectories generated with standard WMC (no background) and in the presence of the harmonic oscillator and linear potentials using the algorithms provided in Appendix A.

\begin{figure}
	\centering
	\includegraphics[width=0.5\textwidth]{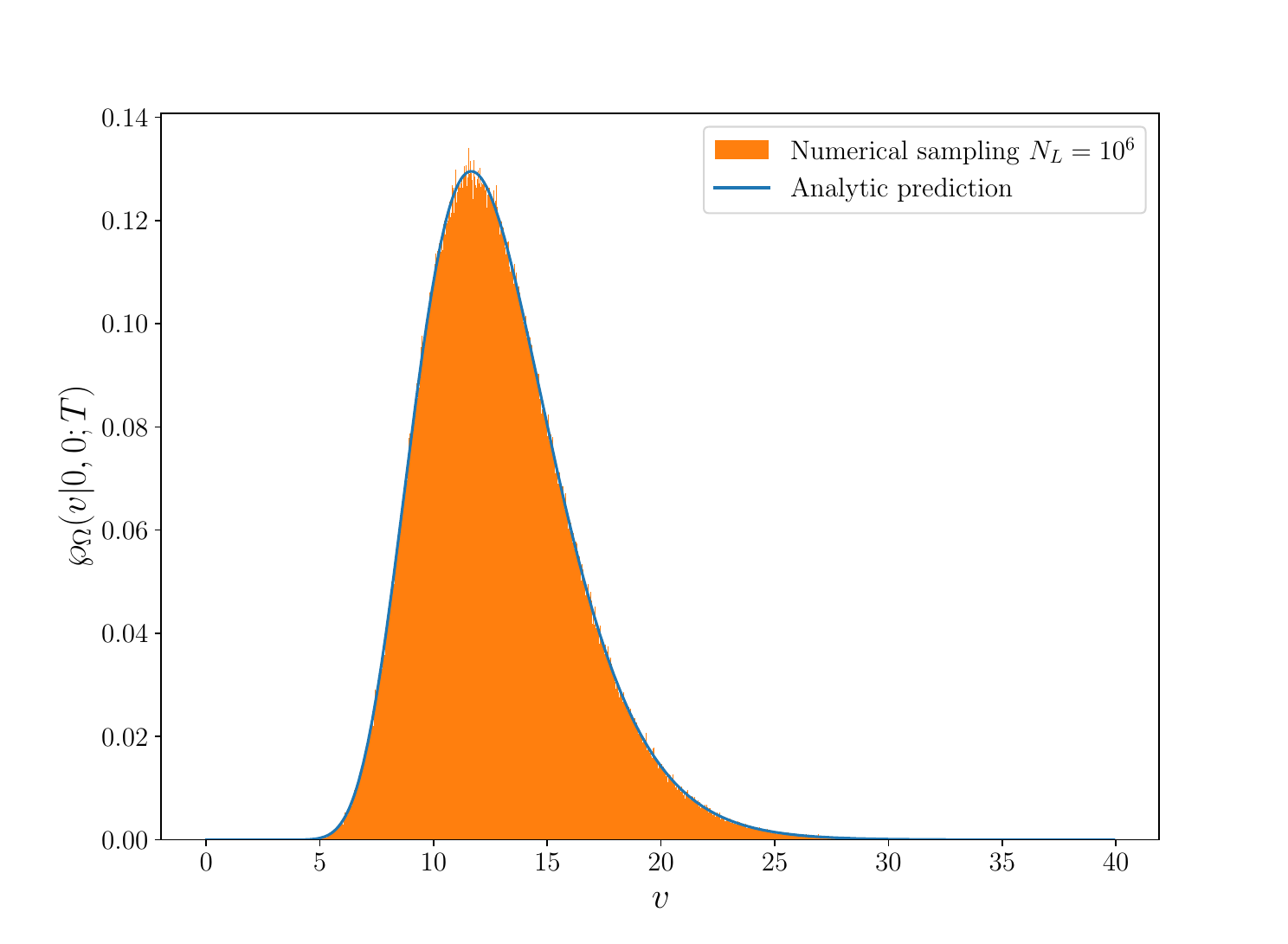}\hfill
	\includegraphics[width=0.5\textwidth]{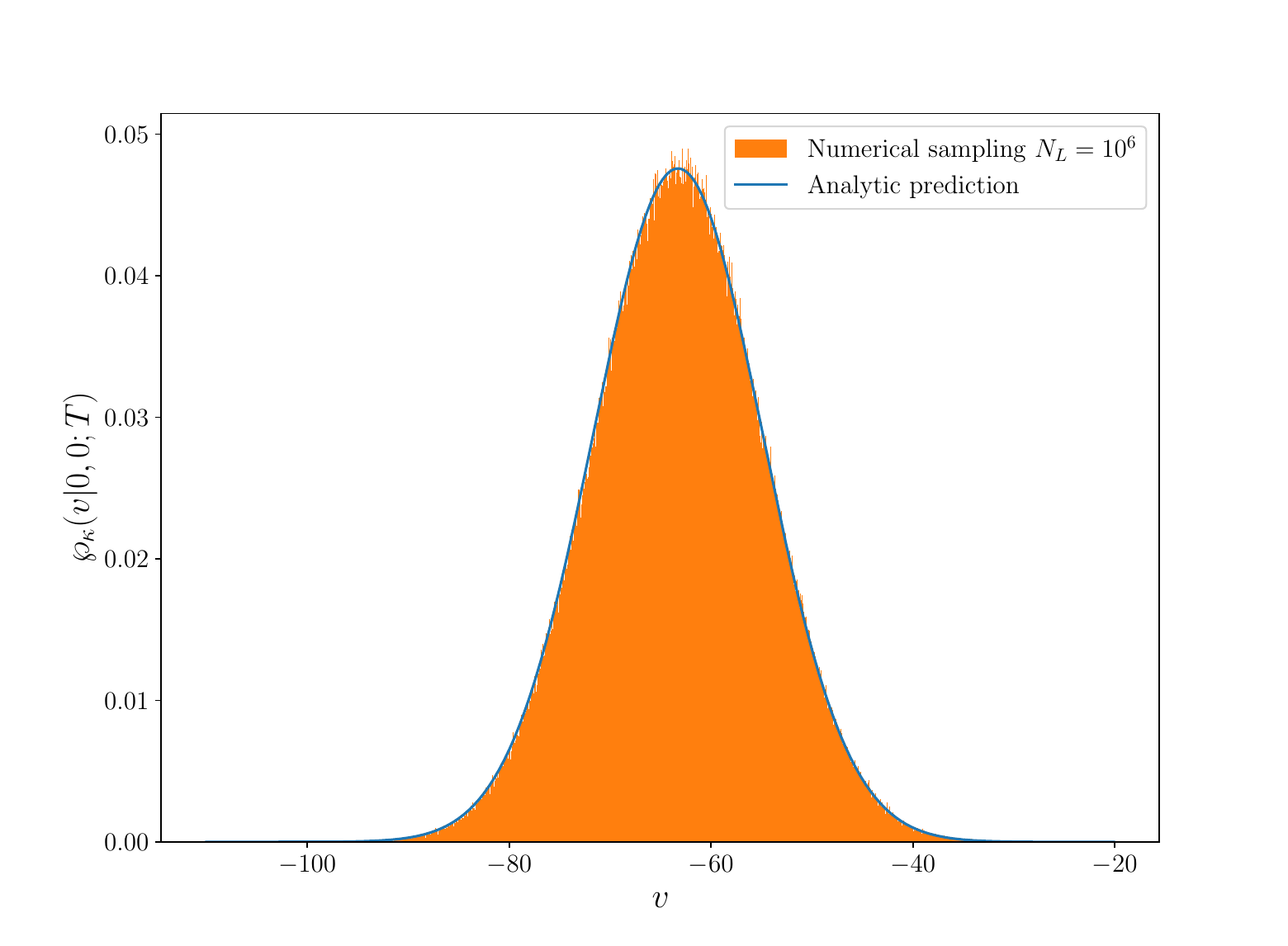}
	\caption{Numerical corroboration of the shifted distributions $\Pv_{\Omega}(v)$ for $\Omega = 0.75$ and $m = 1$, $\omega = 1$, $T = 40$, and $\Pv_{\kappa}(v)$ for $\kappa = 0.45$, and $m = 1$, $k = 0.5$, $T = 15$, using $N_{L} =10^{6} $ trajectories. The blue solid lines represent the analytical results from the Main Text. The small deviations are statistical fluctuations that can be reduced by increasing $N_{L}$ and $N_{P}$.}
	\label{figPvOmegaKappa}
\end{figure}

\begin{figure}
	\centering
	\includegraphics[width=0.5\textwidth]{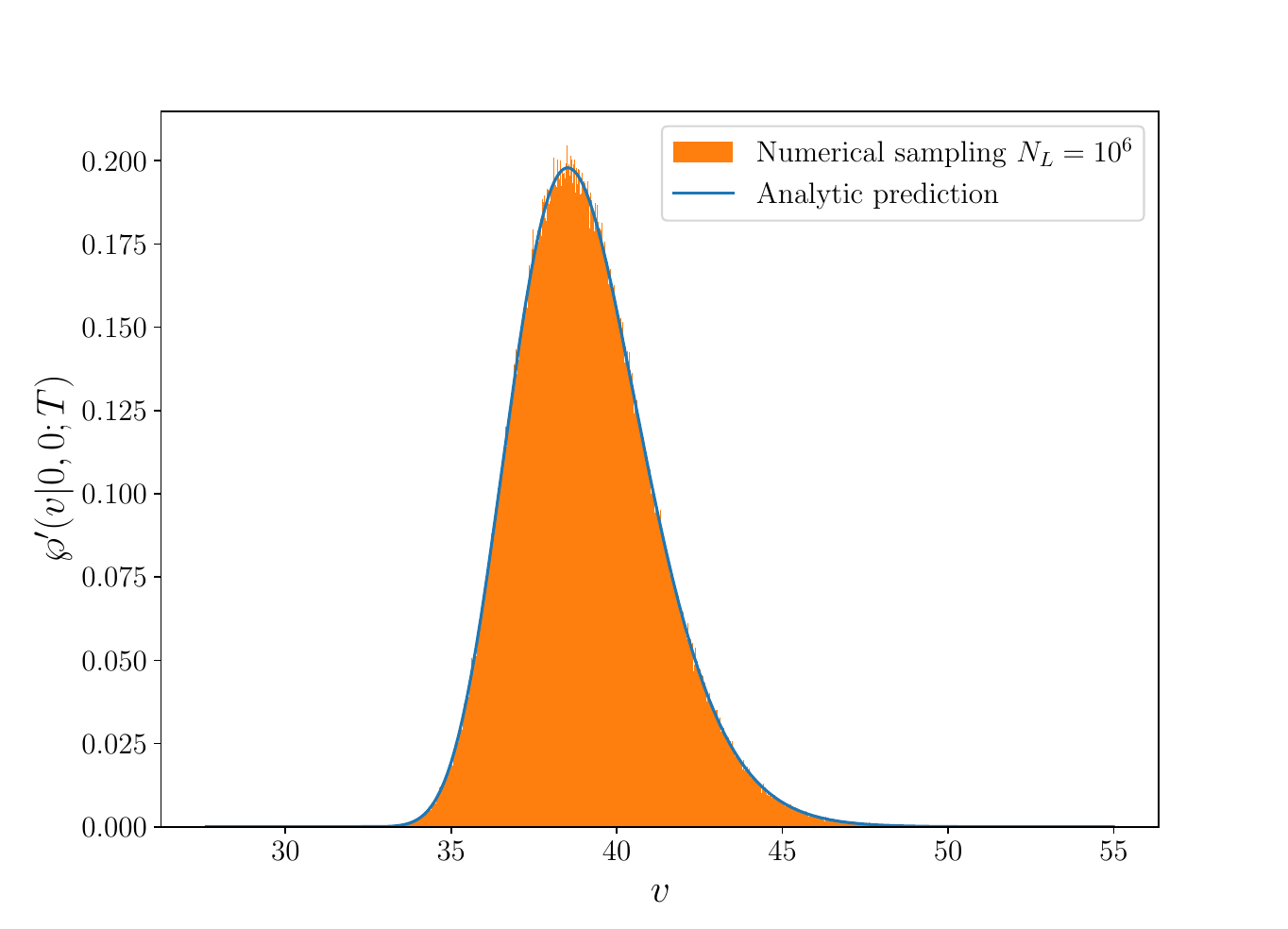}\hfill
	\includegraphics[width=0.5\textwidth]{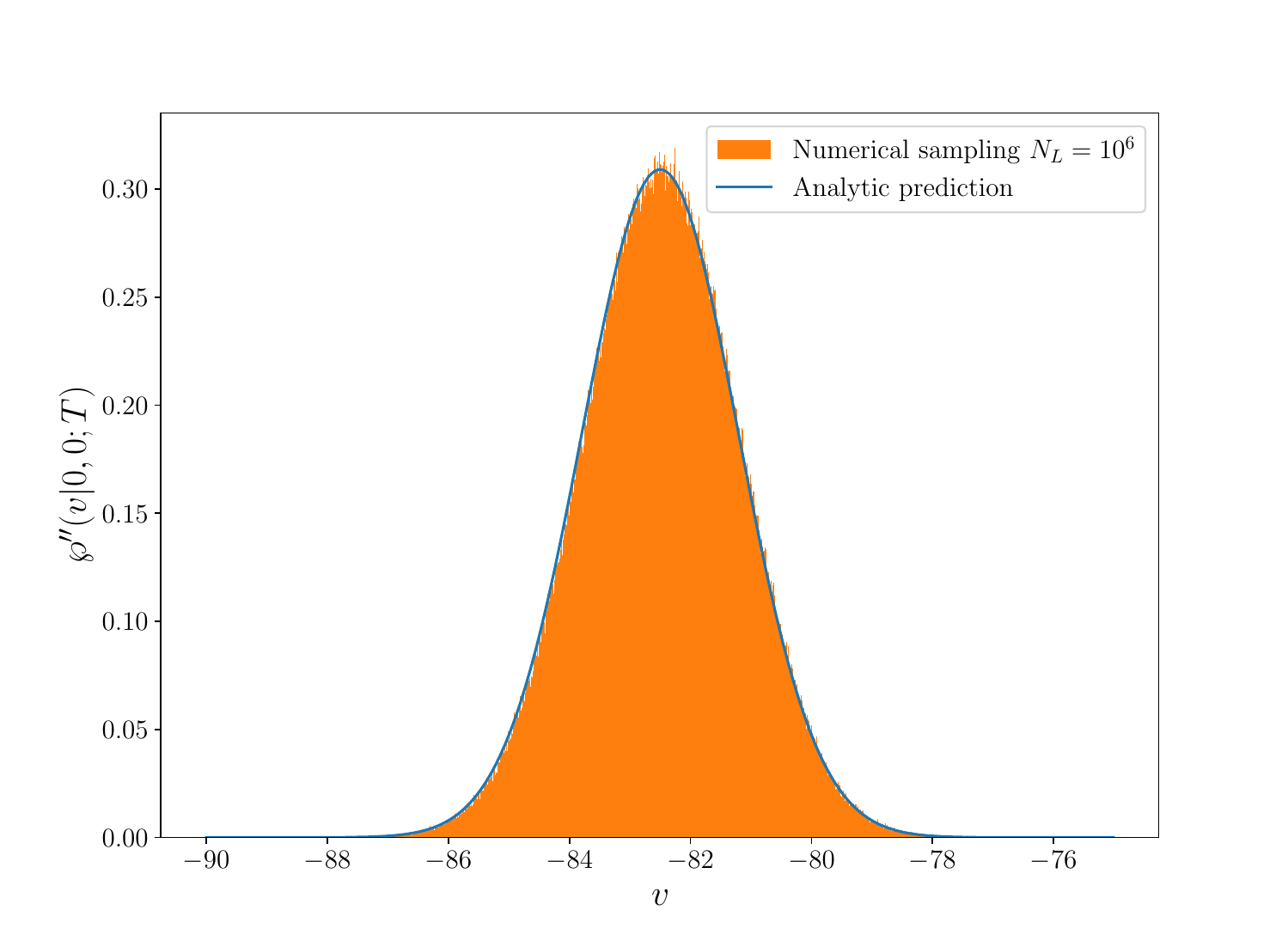}
	\caption{Numerical corroboration of the shifted distributions $\Pvp(v)$, for $\Omega = 0.75$ and $m = 1$, $\omega = 1 $, $T = 80$, and $\Pvpp(v)$, for $\kappa = 0.45$, and $m = 1$, $k = 0.5$, $T = 20$, using $N_{L} = 10^{6}$ trajectories. The blue solid lines represent the analytical results from the Main Text. }
	\label{figPvp}
\end{figure}

The effect of the adjusted spatial growth of the trajectories on values taken by the Wilson line variables can be seen in Figures \ref{figPvOmegaKappa}--\ref{figPvp}, where we show numerical estimation of the distributions on these values against the analytically determined distributions. The numerical estimation is carried out by generating a number, $N_{L}$, of trajectories according to the algorithms provided in Appendix A, which produce a set of values $\{v_{i}\}_{i=1}^{N_{L}}$ of the Wilson line variables of each trajectory. These are then placed into bins to form a density-histogram, which is compared against the functions determined analytically in the Main Text. 

The figures show that the Wilson line variables are providing good samples of the PAP in question, which confirm the correctness of the algorithms for generating trajectories reported in the Main Text. Moreover, comparing these to the ``un-shifted'' PAP (denoted $\Pv(v)$ for both systems), it is apparent that the modified distributions favour \textit{smaller} values of $v$. Now, since the kernel is constructed according to equation (12), trajectories with smaller $v$ dominate its determination. Hence it can be seen that the background is simply concentrating the trajectories to regions that better sample the potential of the systems (in the spirit of importance sampling). As proved in the Main Text, the bias from doing this is removed by the compensation factors which ensure a faithful reconstruction of the propagator, but by having favoured trajectories that better sample the potential, the undersampling problem is mitigated.

\subsubsection{Compensating potentials}
\begin{figure}[t]
	\centering
	\includegraphics[width=0.5\textwidth]{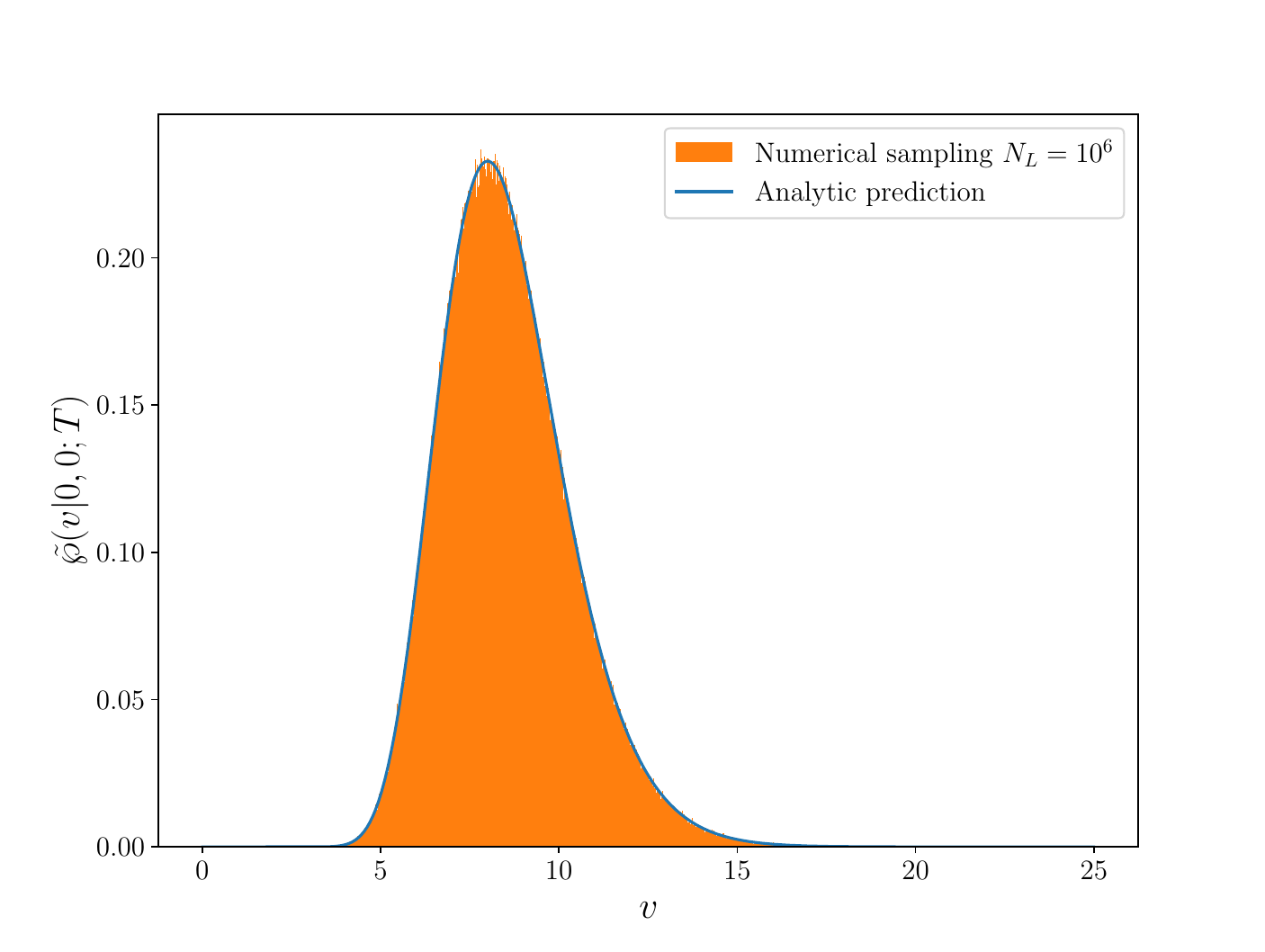}\hfill
	\includegraphics[width=0.5\textwidth]{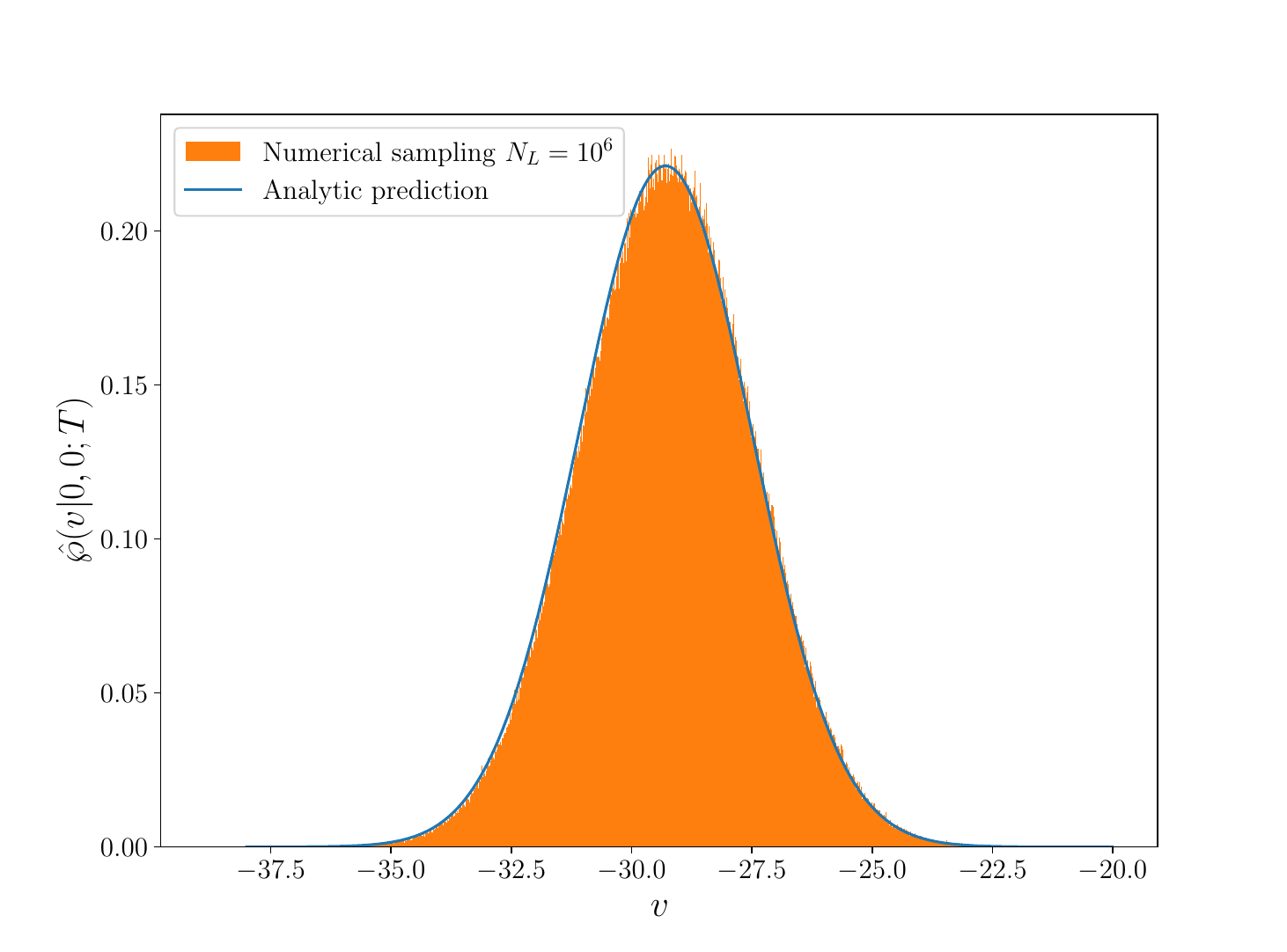}
	\caption{Numerical corroboration of the shifted distributions $\Pvt(v)$ for $\Omega = 0.75$ and $m = 1$, $\omega = 1$, $T = 60$, and $\Pvh(v)$ for $\kappa = 0.45$, and $m = 1$, $k = 0.5$, $T = 25$, using $N_{L} = 10^{6}$ trajectories. The blue solid lines represent the analytical results from the Main Text. }
	\label{figPvTildeHat}
\end{figure}

For systems where the PAP is not known in closed form (e.g. when (12) of the Main Text cannot be evaluated analytically), the compensation factors that ensure the kernel is correctly recovered are unknown. However, the Main Text shows how the bias introduced by the modified algorithms can be removed by subtracting the effects of the background potential directly in the path integral. Figure \ref{figPvTildeHat} demonstrates the numerical sampling of the distributions $\Pvt(v)$ and $\Pvh(v)$ defined in the Main Text, which are distributions on the Wilson line variable for the modified potential.

\section{Applications}
\label{secApA}

We provide additional information on numerical simulation for the absolute value potential, $V_{\kappa}(x) = \kappa |x|$. In this case the propagator is not known in closed form, but its spectral representation can be determined from the Hamiltonian's eigenfunctions and energy eigenvalues (the Green function for this system is given in \cite{Grosche}). The eigenfunctions are Airy ``$\textrm{Ai}$'' functions, defined for $x\geqslant 0$ as $\Psi_{n}(x) = c_{n}\,  \mathrm{Ai}\big( (2mk)^{\frac{1}{3}} \big(x - \frac{E_{n}}{k} \big)\big)$ extended symmetrically (anti-symmetrically) for $n$ even (odd). Their energies are written in terms of zeros of the Airy function and its derivative:
\begin{equation}
	-E_{n} = \Big( \frac{k^{2}}{2m} \Big)^{\frac{1}{3}} \begin{cases}\,  \tilde{\sigma}_{\frac{n+2}{2}}  & n\textrm{ even} \\\, \bar{\sigma}_{\frac{n+1}{2}}  & n\textrm{ odd}  \end{cases}\,\,,
\end{equation}
where the $\bar{\sigma}_{n}$ denote the zeros of $\textrm{Ai}$ and the $\tilde{\sigma}_{n}$ are the zeros of its derivative, $\textrm{Ai}'$ (all lying along the negative real axis). The normalisation constants, $c_{n}$, can be calculated numerically. Truncating the spectral representation provides a good estimate of the kernel for moderate values of propagation time and, in the limit of large times, the ground state, energy $E_{0} = -\Big( \frac{k^{2}}{2m} \Big)^{\frac{1}{3}} \tilde{\sigma}_{1} \approx 1.01879 \Big( \frac{k^{2}}{2m} \Big)^{\frac{1}{3}}$, dominates according to the asymptotic formula in equation (10) of the Main Text. 

\begin{figure}[h]
	\centering
	\includegraphics[width=0.5\textwidth]{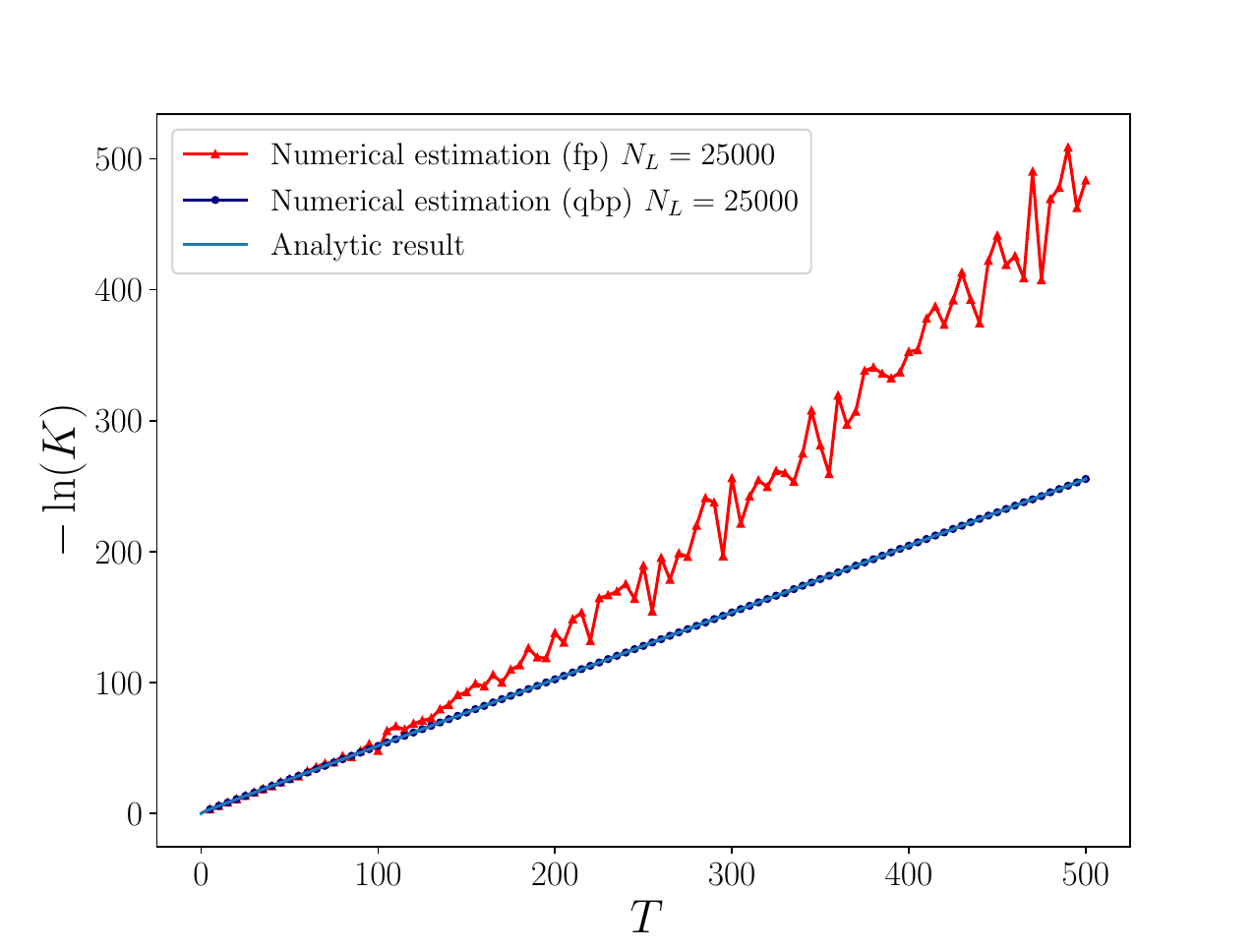}
	\caption{Estimation of the propagator for the absolute value potential, with  $\kappa = 0.5$, and $m = 1$, $\Omega=0.75$, using $N_{L} = 25000$ trajectories generated in a harmonic oscillator background with the potential subtraction method (navy) and $N_p=5000$ points per loop. The blue solid line represent the analytical result based on truncating the spectral representation. The red line with triangles is a estimation using free trajectories.}
	\label{figAbsVal}
\end{figure}

This system was simulated using a harmonic oscillator background ($\Omega = 0.75$), fixing $m = 1$, $\kappa = 0.5$, for which the  ground state energy is $E_{0} = 0.509396\ldots$). This numerical determination of the kernel is shown in Fig \ref{figAbsVal}. Through the linear fit [shown on the graph!] for $T \in [5, 30]$, the energy is estimated to be  $E_{0} = 0.50939336$.

\bibliography{bibSampling2.bib}

\begin{thebibliography}{33}%
\makeatletter
\providecommand \@ifxundefined [1]{%
 \@ifx{#1\undefined}
}%
\providecommand \@ifnum [1]{%
 \ifnum #1\expandafter \@firstoftwo
 \else \expandafter \@secondoftwo
 \fi
}%
\providecommand \@ifx [1]{%
 \ifx #1\expandafter \@firstoftwo
 \else \expandafter \@secondoftwo
 \fi
}%
\providecommand \natexlab [1]{#1}%
\providecommand \enquote  [1]{``#1''}%
\providecommand \bibnamefont  [1]{#1}%
\providecommand \bibfnamefont [1]{#1}%
\providecommand \citenamefont [1]{#1}%
\providecommand \href@noop [0]{\@secondoftwo}%
\providecommand \href [0]{\begingroup \@sanitize@url \@href}%
\providecommand \@href[1]{\@@startlink{#1}\@@href}%
\providecommand \@@href[1]{\endgroup#1\@@endlink}%
\providecommand \@sanitize@url [0]{\catcode `\\12\catcode `\$12\catcode
  `\&12\catcode `\#12\catcode `\^12\catcode `\_12\catcode `\%12\relax}%
\providecommand \@@startlink[1]{}%
\providecommand \@@endlink[0]{}%
\providecommand \url  [0]{\begingroup\@sanitize@url \@url }%
\providecommand \@url [1]{\endgroup\@href {#1}{\urlprefix }}%
\providecommand \urlprefix  [0]{URL }%
\providecommand \Eprint [0]{\href }%
\providecommand \doibase [0]{https://doi.org/}%
\providecommand \selectlanguage [0]{\@gobble}%
\providecommand \bibinfo  [0]{\@secondoftwo}%
\providecommand \bibfield  [0]{\@secondoftwo}%
\providecommand \translation [1]{[#1]}%
\providecommand \BibitemOpen [0]{}%
\providecommand \bibitemStop [0]{}%
\providecommand \bibitemNoStop [0]{.\EOS\space}%
\providecommand \EOS [0]{\spacefactor3000\relax}%
\providecommand \BibitemShut  [1]{\csname bibitem#1\endcsname}%
\let\auto@bib@innerbib\@empty
\bibitem [{\citenamefont {Nieuwenhuis}(1995)}]{TjonThesis}%
  \BibitemOpen
  \bibfield  {author} {\bibinfo {author} {\bibfnamefont {T.}~\bibnamefont
  {Nieuwenhuis}},\ }\emph {\bibinfo {title} {The Feynman-Schwinger
  representation of field theory applied to two-boby bound states}},\
  \href@noop {} {Ph.D. thesis},\ \bibinfo  {school} {Uni. of Utrecht} (\bibinfo
  {year} {1995}),\ \bibinfo {note} {{ISBN 90-393-0890-X}}\BibitemShut {NoStop}%
\bibitem [{\citenamefont {Nieuwenhuis}\ and\ \citenamefont
  {Tjon}(1995)}]{Nieuwenhuis:1995ux}%
  \BibitemOpen
  \bibfield  {author} {\bibinfo {author} {\bibfnamefont {T.}~\bibnamefont
  {Nieuwenhuis}}\ and\ \bibinfo {author} {\bibfnamefont {J.~A.}\ \bibnamefont
  {Tjon}},\ }\bibfield  {title} {\bibinfo {title} {{Bound state solutions of
  scalar QED in (2+1)-dimensions for zero photon mass}},\ }\href
  {https://doi.org/10.1016/0370-2693(95)00753-8} {\bibfield  {journal}
  {\bibinfo  {journal} {Phys. Lett. B}\ }\textbf {\bibinfo {volume} {355}},\
  \bibinfo {pages} {283} (\bibinfo {year} {1995})},\ \Eprint
  {https://arxiv.org/abs/hep-ph/9506346} {arXiv:hep-ph/9506346} \BibitemShut
  {NoStop}%
\bibitem [{\citenamefont {Nieuwenhuis}\ and\ \citenamefont
  {Tjon}(1996)}]{Nieuwenhuis:1996mc}%
  \BibitemOpen
  \bibfield  {author} {\bibinfo {author} {\bibfnamefont {T.}~\bibnamefont
  {Nieuwenhuis}}\ and\ \bibinfo {author} {\bibfnamefont {J.~A.}\ \bibnamefont
  {Tjon}},\ }\bibfield  {title} {\bibinfo {title} {{Nonperturbative study of
  generalized ladder graphs in a phi**2 chi theory}},\ }\href
  {https://doi.org/10.1103/PhysRevLett.77.814} {\bibfield  {journal} {\bibinfo
  {journal} {Phys. Rev. Lett.}\ }\textbf {\bibinfo {volume} {77}},\ \bibinfo
  {pages} {814} (\bibinfo {year} {1996})},\ \Eprint
  {https://arxiv.org/abs/hep-ph/9606403} {arXiv:hep-ph/9606403} \BibitemShut
  {NoStop}%
\bibitem [{\citenamefont {Savkli}\ \emph {et~al.}(1999)\citenamefont {Savkli},
  \citenamefont {Tjon},\ and\ \citenamefont {Gross}}]{Savkli:1999rw}%
  \BibitemOpen
  \bibfield  {author} {\bibinfo {author} {\bibfnamefont {C.}~\bibnamefont
  {Savkli}}, \bibinfo {author} {\bibfnamefont {J.}~\bibnamefont {Tjon}},\ and\
  \bibinfo {author} {\bibfnamefont {F.}~\bibnamefont {Gross}},\ }\bibfield
  {title} {\bibinfo {title} {{Feynman-Schwinger representation approach to
  nonperturbative physics}},\ }\href
  {https://doi.org/10.1103/PhysRevC.60.055210} {\bibfield  {journal} {\bibinfo
  {journal} {Phys. Rev. C}\ }\textbf {\bibinfo {volume} {60}},\ \bibinfo
  {pages} {055210} (\bibinfo {year} {1999})},\ \Eprint
  {https://arxiv.org/abs/hep-ph/9906211} {arXiv:hep-ph/9906211} \BibitemShut
  {NoStop}%
\bibitem [{\citenamefont {Savkli}\ \emph {et~al.}(2002)\citenamefont {Savkli},
  \citenamefont {Gross},\ and\ \citenamefont {Tjon}}]{Savkli:2002fj}%
  \BibitemOpen
  \bibfield  {author} {\bibinfo {author} {\bibfnamefont {C.}~\bibnamefont
  {Savkli}}, \bibinfo {author} {\bibfnamefont {F.}~\bibnamefont {Gross}},\ and\
  \bibinfo {author} {\bibfnamefont {J.}~\bibnamefont {Tjon}},\ }\bibfield
  {title} {\bibinfo {title} {{The Role of interaction vertices in bound state
  calculations}},\ }\href {https://doi.org/10.1016/S0370-2693(02)01362-X}
  {\bibfield  {journal} {\bibinfo  {journal} {Phys. Lett. B}\ }\textbf
  {\bibinfo {volume} {531}},\ \bibinfo {pages} {161} (\bibinfo {year}
  {2002})},\ \Eprint {https://arxiv.org/abs/nucl-th/0202022}
  {arXiv:nucl-th/0202022} \BibitemShut {NoStop}%
\bibitem [{\citenamefont {Gies}\ and\ \citenamefont
  {Langfeld}(2001)}]{Gies:2001zp}%
  \BibitemOpen
  \bibfield  {author} {\bibinfo {author} {\bibfnamefont {H.}~\bibnamefont
  {Gies}}\ and\ \bibinfo {author} {\bibfnamefont {K.}~\bibnamefont
  {Langfeld}},\ }\bibfield  {title} {\bibinfo {title} {{Quantum diffusion of
  magnetic fields in a numerical worldline approach}},\ }\href
  {https://doi.org/10.1016/S0550-3213(01)00377-7} {\bibfield  {journal}
  {\bibinfo  {journal} {Nucl. Phys. B}\ }\textbf {\bibinfo {volume} {613}},\
  \bibinfo {pages} {353} (\bibinfo {year} {2001})},\ \Eprint
  {https://arxiv.org/abs/hep-ph/0102185} {arXiv:hep-ph/0102185} \BibitemShut
  {NoStop}%
\bibitem [{\citenamefont {Gies}\ and\ \citenamefont
  {Langfeld}(2002)}]{Gies:2001tj}%
  \BibitemOpen
  \bibfield  {author} {\bibinfo {author} {\bibfnamefont {H.}~\bibnamefont
  {Gies}}\ and\ \bibinfo {author} {\bibfnamefont {K.}~\bibnamefont
  {Langfeld}},\ }\bibfield  {title} {\bibinfo {title} {{Loops and loop clouds:
  A Numerical approach to the worldline formalism in QED}},\ }\href
  {https://doi.org/10.1142/S0217751X02010388} {\bibfield  {journal} {\bibinfo
  {journal} {Int. J. Mod. Phys. A}\ }\textbf {\bibinfo {volume} {17}},\
  \bibinfo {pages} {966} (\bibinfo {year} {2002})},\ \Eprint
  {https://arxiv.org/abs/hep-ph/0112198} {arXiv:hep-ph/0112198} \BibitemShut
  {NoStop}%
\bibitem [{\citenamefont {Schmidt}\ and\ \citenamefont
  {Stamatescu}(2003)}]{Schmidt:2002mt}%
  \BibitemOpen
  \bibfield  {author} {\bibinfo {author} {\bibfnamefont {M.~G.}\ \bibnamefont
  {Schmidt}}\ and\ \bibinfo {author} {\bibfnamefont {I.-O.}\ \bibnamefont
  {Stamatescu}},\ }\bibfield  {title} {\bibinfo {title} {{Determinant
  calculations using random walk worldline loops}},\ }\href
  {https://doi.org/10.1016/S0920-5632(03)01753-5} {\bibfield  {journal}
  {\bibinfo  {journal} {Nucl. Phys. B Proc. Suppl.}\ }\textbf {\bibinfo
  {volume} {119}},\ \bibinfo {pages} {1030} (\bibinfo {year} {2003})},\ \Eprint
  {https://arxiv.org/abs/hep-lat/0209120} {arXiv:hep-lat/0209120} \BibitemShut
  {NoStop}%
\bibitem [{\citenamefont {Langfeld}\ \emph {et~al.}(2002)\citenamefont
  {Langfeld}, \citenamefont {Moyaerts},\ and\ \citenamefont
  {Gies}}]{Langfeld:2002vy}%
  \BibitemOpen
  \bibfield  {author} {\bibinfo {author} {\bibfnamefont {K.}~\bibnamefont
  {Langfeld}}, \bibinfo {author} {\bibfnamefont {L.}~\bibnamefont {Moyaerts}},\
  and\ \bibinfo {author} {\bibfnamefont {H.}~\bibnamefont {Gies}},\ }\bibfield
  {title} {\bibinfo {title} {{Fermion induced quantum action of vortex
  systems}},\ }\href {https://doi.org/10.1016/S0550-3213(02)00835-0} {\bibfield
   {journal} {\bibinfo  {journal} {Nucl. Phys. B}\ }\textbf {\bibinfo {volume}
  {646}},\ \bibinfo {pages} {158} (\bibinfo {year} {2002})},\ \Eprint
  {https://arxiv.org/abs/hep-th/0205304} {arXiv:hep-th/0205304} \BibitemShut
  {NoStop}%
\bibitem [{\citenamefont {Gies}\ \emph {et~al.}(2003)\citenamefont {Gies},
  \citenamefont {Langfeld},\ and\ \citenamefont {Moyaerts}}]{Gies:2003cv}%
  \BibitemOpen
  \bibfield  {author} {\bibinfo {author} {\bibfnamefont {H.}~\bibnamefont
  {Gies}}, \bibinfo {author} {\bibfnamefont {K.}~\bibnamefont {Langfeld}},\
  and\ \bibinfo {author} {\bibfnamefont {L.}~\bibnamefont {Moyaerts}},\
  }\bibfield  {title} {\bibinfo {title} {{Casimir effect on the worldline}},\
  }\href {https://doi.org/10.1088/1126-6708/2003/06/018} {\bibfield  {journal}
  {\bibinfo  {journal} {JHEP}\ }\textbf {\bibinfo {volume} {06}},\ \bibinfo
  {pages} {018}},\ \Eprint {https://arxiv.org/abs/hep-th/0303264}
  {arXiv:hep-th/0303264} \BibitemShut {NoStop}%
\bibitem [{\citenamefont {Gies}\ \emph {et~al.}(2005)\citenamefont {Gies},
  \citenamefont {Sanchez-Guillen},\ and\ \citenamefont
  {Vazquez}}]{Gies:2005sb}%
  \BibitemOpen
  \bibfield  {author} {\bibinfo {author} {\bibfnamefont {H.}~\bibnamefont
  {Gies}}, \bibinfo {author} {\bibfnamefont {J.}~\bibnamefont
  {Sanchez-Guillen}},\ and\ \bibinfo {author} {\bibfnamefont {R.~A.}\
  \bibnamefont {Vazquez}},\ }\bibfield  {title} {\bibinfo {title} {{Quantum
  effective actions from nonperturbative worldline dynamics}},\ }\href
  {https://doi.org/10.1088/1126-6708/2005/08/067} {\bibfield  {journal}
  {\bibinfo  {journal} {JHEP}\ }\textbf {\bibinfo {volume} {08}},\ \bibinfo
  {pages} {067}},\ \Eprint {https://arxiv.org/abs/hep-th/0505275}
  {arXiv:hep-th/0505275} \BibitemShut {NoStop}%
\bibitem [{\citenamefont {Langfeld}\ \emph {et~al.}(2007)\citenamefont
  {Langfeld}, \citenamefont {Dunne}, \citenamefont {Gies},\ and\ \citenamefont
  {Klingmuller}}]{Langfeld:2007wh}%
  \BibitemOpen
  \bibfield  {author} {\bibinfo {author} {\bibfnamefont {K.}~\bibnamefont
  {Langfeld}}, \bibinfo {author} {\bibfnamefont {G.}~\bibnamefont {Dunne}},
  \bibinfo {author} {\bibfnamefont {H.}~\bibnamefont {Gies}},\ and\ \bibinfo
  {author} {\bibfnamefont {K.}~\bibnamefont {Klingmuller}},\ }\bibfield
  {title} {\bibinfo {title} {{Worldline Approach to Chiral Fermions}},\ }\href
  {https://doi.org/10.22323/1.042.0202} {\bibfield  {journal} {\bibinfo
  {journal} {PoS}\ }\textbf {\bibinfo {volume} {LATTICE2007}},\ \bibinfo
  {pages} {202} (\bibinfo {year} {2007})},\ \Eprint
  {https://arxiv.org/abs/0709.4595} {arXiv:0709.4595 [hep-lat]} \BibitemShut
  {NoStop}%
\bibitem [{\citenamefont {Strassler}(1992)}]{Strass1}%
  \BibitemOpen
  \bibfield  {author} {\bibinfo {author} {\bibfnamefont {M.~J.}\ \bibnamefont
  {Strassler}},\ }\bibfield  {title} {\bibinfo {title} {Field theory without
  {F}eynman diagrams: One loop effective actions},\ }\href@noop {} {\bibfield
  {journal} {\bibinfo  {journal} {Nucl. Phys. B}\ }\textbf {\bibinfo {volume}
  {385}},\ \bibinfo {pages} {145} (\bibinfo {year} {1992})}\BibitemShut
  {NoStop}%
\bibitem [{\citenamefont {Schmidt}\ and\ \citenamefont
  {Schubert}(1998)}]{SchmidtRev}%
  \BibitemOpen
  \bibfield  {author} {\bibinfo {author} {\bibfnamefont {M.~G.}\ \bibnamefont
  {Schmidt}}\ and\ \bibinfo {author} {\bibfnamefont {C.}~\bibnamefont
  {Schubert}},\ }\bibfield  {title} {\bibinfo {title} {{Worldline path
  integrals as a calculational tool in quantum field theory}},\ }in\ \href@noop
  {} {\emph {\bibinfo {booktitle} {{6th International Conference on Path
  Integrals from PeV to TeV}}}}\ (\bibinfo {year} {1998})\ \Eprint
  {https://arxiv.org/abs/hep-th/9810161} {arXiv:hep-th/9810161} \BibitemShut
  {NoStop}%
\bibitem [{\citenamefont {Schubert}(2001)}]{ChrisRev}%
  \BibitemOpen
  \bibfield  {author} {\bibinfo {author} {\bibfnamefont {C.}~\bibnamefont
  {Schubert}},\ }\bibfield  {title} {\bibinfo {title} {{Perturbative quantum
  field theory in the string inspired formalism}},\ }\href
  {https://doi.org/10.1016/S0370-1573(01)00013-8} {\bibfield  {journal}
  {\bibinfo  {journal} {Phys. Rept.}\ }\textbf {\bibinfo {volume} {355}},\
  \bibinfo {pages} {73} (\bibinfo {year} {2001})},\ \Eprint
  {https://arxiv.org/abs/hep-th/0101036} {arXiv:hep-th/0101036} \BibitemShut
  {NoStop}%
\bibitem [{\citenamefont {Edwards}\ and\ \citenamefont
  {Schubert}(2019)}]{UsRep}%
  \BibitemOpen
  \bibfield  {author} {\bibinfo {author} {\bibfnamefont {J.~P.}\ \bibnamefont
  {Edwards}}\ and\ \bibinfo {author} {\bibfnamefont {C.}~\bibnamefont
  {Schubert}},\ }\bibfield  {title} {\bibinfo {title} {{Quantum mechanical path
  integrals in the first quantised approach to quantum field theory}}\
  }(\bibinfo {year} {2019})\ \Eprint {https://arxiv.org/abs/1912.10004}
  {arXiv:1912.10004 [hep-th]} \BibitemShut {NoStop}%
\bibitem [{\citenamefont {Gies}\ and\ \citenamefont
  {Klingmuller}(2005)}]{Gies:2005bz}%
  \BibitemOpen
  \bibfield  {author} {\bibinfo {author} {\bibfnamefont {H.}~\bibnamefont
  {Gies}}\ and\ \bibinfo {author} {\bibfnamefont {K.}~\bibnamefont
  {Klingmuller}},\ }\bibfield  {title} {\bibinfo {title} {{Pair production in
  inhomogeneous fields}},\ }\href {https://doi.org/10.1103/PhysRevD.72.065001}
  {\bibfield  {journal} {\bibinfo  {journal} {Phys. Rev. D}\ }\textbf {\bibinfo
  {volume} {72}},\ \bibinfo {pages} {065001} (\bibinfo {year}
  {2005})}\BibitemShut {NoStop}%
\bibitem [{\citenamefont {Gies}\ and\ \citenamefont
  {Klingmuller}(2006{\natexlab{a}})}]{Gies:2005ym}%
  \BibitemOpen
  \bibfield  {author} {\bibinfo {author} {\bibfnamefont {H.}~\bibnamefont
  {Gies}}\ and\ \bibinfo {author} {\bibfnamefont {K.}~\bibnamefont
  {Klingmuller}},\ }\bibfield  {title} {\bibinfo {title} {{Quantum energies
  with worldline numerics}},\ }\href
  {https://doi.org/10.1088/0305-4470/39/21/S36} {\bibfield  {journal} {\bibinfo
   {journal} {J. Phys. A}\ }\textbf {\bibinfo {volume} {39}},\ \bibinfo {pages}
  {6415} (\bibinfo {year} {2006}{\natexlab{a}})},\ \Eprint
  {https://arxiv.org/abs/hep-th/0511092} {arXiv:hep-th/0511092} \BibitemShut
  {NoStop}%
\bibitem [{\citenamefont {Moyaerts}\ \emph {et~al.}(2003)\citenamefont
  {Moyaerts}, \citenamefont {Langfeld},\ and\ \citenamefont
  {Gies}}]{Moyaerts:2003ts}%
  \BibitemOpen
  \bibfield  {author} {\bibinfo {author} {\bibfnamefont {L.}~\bibnamefont
  {Moyaerts}}, \bibinfo {author} {\bibfnamefont {K.}~\bibnamefont {Langfeld}},\
  and\ \bibinfo {author} {\bibfnamefont {H.}~\bibnamefont {Gies}},\ }\bibfield
  {title} {\bibinfo {title} {{Worldline approach to the Casimir effect}},\ }in\
  \href@noop {} {\emph {\bibinfo {booktitle} {{6th Workshop on Quantum Field
  Theory under the Influence of External Conditions}}}}\ (\bibinfo {year}
  {2003})\ pp.\ \bibinfo {pages} {203--211},\ \Eprint
  {https://arxiv.org/abs/hep-th/0311168} {arXiv:hep-th/0311168} \BibitemShut
  {NoStop}%
\bibitem [{\citenamefont {Gies}\ and\ \citenamefont
  {Klingmuller}(2006{\natexlab{b}})}]{Gies:2006cq}%
  \BibitemOpen
  \bibfield  {author} {\bibinfo {author} {\bibfnamefont {H.}~\bibnamefont
  {Gies}}\ and\ \bibinfo {author} {\bibfnamefont {K.}~\bibnamefont
  {Klingmuller}},\ }\bibfield  {title} {\bibinfo {title} {{Worldline algorithms
  for Casimir configurations}},\ }\href
  {https://doi.org/10.1103/PhysRevD.74.045002} {\bibfield  {journal} {\bibinfo
  {journal} {Phys. Rev. D}\ }\textbf {\bibinfo {volume} {74}},\ \bibinfo
  {pages} {045002} (\bibinfo {year} {2006}{\natexlab{b}})},\ \Eprint
  {https://arxiv.org/abs/quant-ph/0605141} {arXiv:quant-ph/0605141}
  \BibitemShut {NoStop}%
\bibitem [{\citenamefont {Gies}\ and\ \citenamefont
  {Klingmuller}(2006{\natexlab{c}})}]{Gies:2006bt}%
  \BibitemOpen
  \bibfield  {author} {\bibinfo {author} {\bibfnamefont {H.}~\bibnamefont
  {Gies}}\ and\ \bibinfo {author} {\bibfnamefont {K.}~\bibnamefont
  {Klingmuller}},\ }\bibfield  {title} {\bibinfo {title} {{Casimir effect for
  curved geometries: PFA validity limits}},\ }\href
  {https://doi.org/10.1103/PhysRevLett.96.220401} {\bibfield  {journal}
  {\bibinfo  {journal} {Phys. Rev. Lett.}\ }\textbf {\bibinfo {volume} {96}},\
  \bibinfo {pages} {220401} (\bibinfo {year} {2006}{\natexlab{c}})},\ \Eprint
  {https://arxiv.org/abs/quant-ph/0601094} {arXiv:quant-ph/0601094}
  \BibitemShut {NoStop}%
\bibitem [{\citenamefont {Weber}\ and\ \citenamefont
  {Gies}(2009)}]{Weber:2009dp}%
  \BibitemOpen
  \bibfield  {author} {\bibinfo {author} {\bibfnamefont {A.}~\bibnamefont
  {Weber}}\ and\ \bibinfo {author} {\bibfnamefont {H.}~\bibnamefont {Gies}},\
  }\bibfield  {title} {\bibinfo {title} {{Interplay between geometry and
  temperature for inclined Casimir plates}},\ }\href
  {https://doi.org/10.1103/PhysRevD.80.065033} {\bibfield  {journal} {\bibinfo
  {journal} {Phys. Rev. D}\ }\textbf {\bibinfo {volume} {80}},\ \bibinfo
  {pages} {065033} (\bibinfo {year} {2009})},\ \Eprint
  {https://arxiv.org/abs/0906.2313} {arXiv:0906.2313 [hep-th]} \BibitemShut
  {NoStop}%
\bibitem [{\citenamefont {Schaden}(2009)}]{PhysRevLett.102.060402}%
  \BibitemOpen
  \bibfield  {author} {\bibinfo {author} {\bibfnamefont {M.}~\bibnamefont
  {Schaden}},\ }\bibfield  {title} {\bibinfo {title} {Dependence of the
  direction of the casimir force on the shape of the boundary},\ }\href
  {https://doi.org/10.1103/PhysRevLett.102.060402} {\bibfield  {journal}
  {\bibinfo  {journal} {Phys. Rev. Lett.}\ }\textbf {\bibinfo {volume} {102}},\
  \bibinfo {pages} {060402} (\bibinfo {year} {2009})}\BibitemShut {NoStop}%
\bibitem [{\citenamefont {Schafer}\ \emph {et~al.}(2016)\citenamefont
  {Schafer}, \citenamefont {Huet},\ and\ \citenamefont {Gies}}]{Idrish}%
  \BibitemOpen
  \bibfield  {author} {\bibinfo {author} {\bibfnamefont {M.}~\bibnamefont
  {Schafer}}, \bibinfo {author} {\bibfnamefont {I.}~\bibnamefont {Huet}},\ and\
  \bibinfo {author} {\bibfnamefont {H.}~\bibnamefont {Gies}},\ }\bibfield
  {title} {\bibinfo {title} {{Worldline Numerics for Energy-Momentum Tensors in
  Casimir Geometries}},\ }\href
  {https://doi.org/10.1088/1751-8113/49/13/135402} {\bibfield  {journal}
  {\bibinfo  {journal} {J. Phys. A}\ }\textbf {\bibinfo {volume} {49}},\
  \bibinfo {pages} {135402} (\bibinfo {year} {2016})},\ \Eprint
  {https://arxiv.org/abs/1509.03509} {arXiv:1509.03509 [hep-th]} \BibitemShut
  {NoStop}%
\bibitem [{\citenamefont {Schafer}\ \emph {et~al.}(2012)\citenamefont
  {Schafer}, \citenamefont {Huet},\ and\ \citenamefont {Gies}}]{Idrish2}%
  \BibitemOpen
  \bibfield  {author} {\bibinfo {author} {\bibfnamefont {M.}~\bibnamefont
  {Schafer}}, \bibinfo {author} {\bibfnamefont {I.}~\bibnamefont {Huet}},\ and\
  \bibinfo {author} {\bibfnamefont {H.}~\bibnamefont {Gies}},\ }\bibfield
  {title} {\bibinfo {title} {{Energy-Momentum Tensors with Worldline
  Numerics}},\ }\href {https://doi.org/10.1142/S2010194512007647} {\bibfield
  {journal} {\bibinfo  {journal} {Int. J. Mod. Phys. Conf. Ser.}\ }\textbf
  {\bibinfo {volume} {14}},\ \bibinfo {pages} {511} (\bibinfo {year} {2012})},\
  \Eprint {https://arxiv.org/abs/1112.0469} {arXiv:1112.0469 [hep-th]}
  \BibitemShut {NoStop}%
\bibitem [{\citenamefont {Edwards}\ \emph {et~al.}(2019)\citenamefont
  {Edwards}, \citenamefont {Gerber}, \citenamefont {Schubert}, \citenamefont
  {Trejo}, \citenamefont {Tsiftsi},\ and\ \citenamefont
  {Weber}}]{UsMonteCarlo}%
  \BibitemOpen
  \bibfield  {author} {\bibinfo {author} {\bibfnamefont {J.~P.}\ \bibnamefont
  {Edwards}}, \bibinfo {author} {\bibfnamefont {U.}~\bibnamefont {Gerber}},
  \bibinfo {author} {\bibfnamefont {C.}~\bibnamefont {Schubert}}, \bibinfo
  {author} {\bibfnamefont {M.~A.}\ \bibnamefont {Trejo}}, \bibinfo {author}
  {\bibfnamefont {T.}~\bibnamefont {Tsiftsi}},\ and\ \bibinfo {author}
  {\bibfnamefont {A.}~\bibnamefont {Weber}},\ }\bibfield  {title} {\bibinfo
  {title} {{Applications of the worldline Monte Carlo formalism in quantum
  mechanics}},\ }\href {https://doi.org/10.1016/j.aop.2019.167966} {\bibfield
  {journal} {\bibinfo  {journal} {Annals Phys.}\ }\textbf {\bibinfo {volume}
  {411}},\ \bibinfo {pages} {167966} (\bibinfo {year} {2019})},\ \Eprint
  {https://arxiv.org/abs/1903.00536} {arXiv:1903.00536 [quant-ph]} \BibitemShut
  {NoStop}%
\bibitem [{\citenamefont {Franchino-Vi\~nas}\ and\ \citenamefont
  {Gies}(2019)}]{Franchino-Vinas:2019udt}%
  \BibitemOpen
  \bibfield  {author} {\bibinfo {author} {\bibfnamefont {S.}~\bibnamefont
  {Franchino-Vi\~nas}}\ and\ \bibinfo {author} {\bibfnamefont {H.}~\bibnamefont
  {Gies}},\ }\bibfield  {title} {\bibinfo {title} {{Propagator from
  Nonperturbative Worldline Dynamics}},\ }\href
  {https://doi.org/10.1103/PhysRevD.100.105020} {\bibfield  {journal} {\bibinfo
   {journal} {Phys. Rev. D}\ }\textbf {\bibinfo {volume} {100}},\ \bibinfo
  {pages} {105020} (\bibinfo {year} {2019})},\ \Eprint
  {https://arxiv.org/abs/1908.04532} {arXiv:1908.04532 [hep-th]} \BibitemShut
  {NoStop}%
\bibitem [{\citenamefont {Corradini}\ and\ \citenamefont
  {Muratori}(2020)}]{Corradini:2020tgk}%
  \BibitemOpen
  \bibfield  {author} {\bibinfo {author} {\bibfnamefont {O.}~\bibnamefont
  {Corradini}}\ and\ \bibinfo {author} {\bibfnamefont {M.}~\bibnamefont
  {Muratori}},\ }\bibfield  {title} {\bibinfo {title} {{A Monte Carlo Approach
  to the Worldline Formalism in Curved Space}},\ }\href
  {https://doi.org/10.1007/JHEP11(2020)169} {\bibfield  {journal} {\bibinfo
  {journal} {JHEP}\ }\textbf {\bibinfo {volume} {11}},\ \bibinfo {pages}
  {169}},\ \Eprint {https://arxiv.org/abs/2006.02911} {arXiv:2006.02911
  [hep-th]} \BibitemShut {NoStop}%
\bibitem [{\citenamefont {Dunne}\ \emph {et~al.}(2009)\citenamefont {Dunne},
  \citenamefont {Gies}, \citenamefont {Klingm{\"u}ller},\ and\ \citenamefont
  {Langfeld}}]{dunne2009worldline}%
  \BibitemOpen
  \bibfield  {author} {\bibinfo {author} {\bibfnamefont {G.}~\bibnamefont
  {Dunne}}, \bibinfo {author} {\bibfnamefont {H.}~\bibnamefont {Gies}},
  \bibinfo {author} {\bibfnamefont {K.}~\bibnamefont {Klingm{\"u}ller}},\ and\
  \bibinfo {author} {\bibfnamefont {K.}~\bibnamefont {Langfeld}},\ }\bibfield
  {title} {\bibinfo {title} {Worldline monte carlo for fermion models at large
  nf},\ }\href@noop {} {\bibfield  {journal} {\bibinfo  {journal} {Journal of
  High Energy Physics}\ }\textbf {\bibinfo {volume} {2009}},\ \bibinfo {pages}
  {010} (\bibinfo {year} {2009})}\BibitemShut {NoStop}%
\bibitem [{\citenamefont {Edwards}\ \emph {et~al.}(2018)\citenamefont
  {Edwards}, \citenamefont {Gerber}, \citenamefont {Schubert}, \citenamefont
  {Trejo},\ and\ \citenamefont {Weber}}]{PvHz}%
  \BibitemOpen
  \bibfield  {author} {\bibinfo {author} {\bibfnamefont {J.~P.}\ \bibnamefont
  {Edwards}}, \bibinfo {author} {\bibfnamefont {U.}~\bibnamefont {Gerber}},
  \bibinfo {author} {\bibfnamefont {C.}~\bibnamefont {Schubert}}, \bibinfo
  {author} {\bibfnamefont {M.~A.}\ \bibnamefont {Trejo}},\ and\ \bibinfo
  {author} {\bibfnamefont {A.}~\bibnamefont {Weber}},\ }\bibfield  {title}
  {\bibinfo {title} {{Integral transforms of the quantum mechanical path
  integral: hit function and path averaged potential}},\ }\href
  {https://doi.org/10.1103/PhysRevE.97.042114} {\bibfield  {journal} {\bibinfo
  {journal} {Phys. Rev. E}\ }\textbf {\bibinfo {volume} {97}},\ \bibinfo
  {pages} {042114} (\bibinfo {year} {2018})},\ \Eprint
  {https://arxiv.org/abs/1709.04984} {arXiv:1709.04984 [quant-ph]} \BibitemShut
  {NoStop}%
\bibitem [{\citenamefont {Edwards}\ \emph {et~al.}(2022)\citenamefont
  {Edwards}, \citenamefont {Gonz\'alez-Dom\'\i{}nguez}, \citenamefont {Huet},\
  and\ \citenamefont {Trejo}}]{nHit}%
  \BibitemOpen
  \bibfield  {author} {\bibinfo {author} {\bibfnamefont {J.~P.}\ \bibnamefont
  {Edwards}}, \bibinfo {author} {\bibfnamefont {V.~A.}\ \bibnamefont
  {Gonz\'alez-Dom\'\i{}nguez}}, \bibinfo {author} {\bibfnamefont
  {I.}~\bibnamefont {Huet}},\ and\ \bibinfo {author} {\bibfnamefont {M.~A.}\
  \bibnamefont {Trejo}},\ }\bibfield  {title} {\bibinfo {title} {{Resummation
  for quantum propagators in bounded spaces}},\ }\href
  {https://doi.org/10.1103/PhysRevE.105.064132} {\bibfield  {journal} {\bibinfo
   {journal} {Phys. Rev. E}\ }\textbf {\bibinfo {volume} {105}},\ \bibinfo
  {pages} {064132} (\bibinfo {year} {2022})},\ \Eprint
  {https://arxiv.org/abs/2110.04969} {arXiv:2110.04969 [quant-ph]} \BibitemShut
  {NoStop}%
\bibitem [{\citenamefont {Ahmadiniaz}\ \emph {et~al.}(2022)\citenamefont
  {Ahmadiniaz}, \citenamefont {Franchino-Vi\~nas}, \citenamefont {Manzo},\ and\
  \citenamefont {Mazzitelli}}]{SebaNaser}%
  \BibitemOpen
  \bibfield  {author} {\bibinfo {author} {\bibfnamefont {N.}~\bibnamefont
  {Ahmadiniaz}}, \bibinfo {author} {\bibfnamefont {S.~A.}\ \bibnamefont
  {Franchino-Vi\~nas}}, \bibinfo {author} {\bibfnamefont {L.}~\bibnamefont
  {Manzo}},\ and\ \bibinfo {author} {\bibfnamefont {F.~D.}\ \bibnamefont
  {Mazzitelli}},\ }\bibfield  {title} {\bibinfo {title} {{Local Neumann
  semitransparent layers: resummation, pair production and duality}},\
  }\href@noop {} {\  (\bibinfo {year} {2022})},\ \Eprint
  {https://arxiv.org/abs/2208.07383} {arXiv:2208.07383 [hep-th]} \BibitemShut
  {NoStop}%
\bibitem [{\citenamefont {Grosche}\ and\ \citenamefont
  {Steiner}(1998)}]{Grosche}%
  \BibitemOpen
  \bibfield  {author} {\bibinfo {author} {\bibfnamefont {C.}~\bibnamefont
  {Grosche}}\ and\ \bibinfo {author} {\bibfnamefont {F.}~\bibnamefont
  {Steiner}},\ }\href@noop {} {\emph {\bibinfo {title} {{Handbook of Feynman
  Path Integrals}}}},\ Vol.\ \bibinfo {volume} {145}\ (\bibinfo {year}
  {1998})\BibitemShut {NoStop}%
\end{thebibliography}%
\end{document}